\documentclass[preprint2]{aastex}

\newcommand{\etal}{\mbox{et\ al.}}

\newcommand{\aanda}  {A\&A\nolinebreak }

\usepackage{rotating}
\usepackage{amssymb}
\usepackage{longtable}

\shorttitle{GRB afterglows as probes of environment}
\shortauthors{Starling et al.}

\begin{document}

\title
{Gamma-Ray Burst afterglows as probes of environment and blastwave physics I:
  absorption by host galaxy gas and dust}

\author{R. L. C. Starling\altaffilmark{1,2,*}, R. A. M. J. Wijers\altaffilmark{1}, K. Wiersema\altaffilmark{1},
  E. Rol\altaffilmark{2}, P. A. Curran\altaffilmark{1},
  C. Kouveliotou\altaffilmark{3}, A. J. van der Horst\altaffilmark{1} and
  M. H. M. Heemskerk\altaffilmark{1} }
\altaffiltext{1}{Astronomical Institute `Anton Pannekoek', University of Amsterdam, Kruislaan 403, 1098 SJ Amsterdam, The Netherlands}
\altaffiltext{2}{Department of Physics and Astronomy, University of Leicester, University Road, Leicester LE1 7RH, UK}
\altaffiltext{3}{NASA Marshall Space Flight Center, NSSTC, VP-62, 320 Sparkman Drive, Huntsville, AL 35805, USA}
\altaffiltext{*}{email rlcs1@star.le.ac.uk}


\begin{abstract}
We use a new approach to obtain limits on the absorbing columns towards an initial
sample of 10 long Gamma-Ray Bursts observed with {\it BeppoSAX} and selected on
the basis of their good optical and nIR coverage, from simultaneous
fits to nIR, optical and X-ray afterglow data, in count space and including
the effects of metallicity. In no cases is a MW-like extinction preferred,
when testing MW, LMC and SMC extinction laws. The 2175\AA\ bump would in
principle be detectable in all these afterglows, but is not present in the
data. An SMC-like gas-to-dust ratio or lower value can be ruled out for 4 of
the hosts analysed here (assuming SMC metallicity and extinction law) whilst
the remainder of the sample have too large an error to discriminate. We
provide a more accurate estimate of the line-of-sight extinction and
improve upon the uncertainties for the majority of the extinction measurements made in previous studies of this sample. We discuss this method to determine extinction values in comparison
with the most commonly employed existing methods.\\
\end{abstract}


\section{Introduction}
The accurate localisation of Gamma-Ray Bursts (GRBs) through their optical and
X-ray afterglows has enabled detailed studies of their environments. Selection by the unobscured gamma-ray flash alone has allowed the
discovery of a unique sample of galaxies which span an enormously wide range
of redshifts from $z$ $\sim$ 0.009 (GRB\,980425, e.g. Tinney et
al. 1998; Galama et al. 1998a) to 6.3 (GRB\,050904, Kawai et al. 2005, 2006).
The subset of long-duration ($>$ 2 s) GRBs are almost certainly caused by the collapse of
certain massive stars to black holes, confirmed by observations of supernova
components in the late-time afterglows of a number of long-GRBs (Woosley \&
Bloom 2006; Kaneko et al. 2006) and by the observed location of GRBs in
UV-bright regions within their host galaxies (Bloom et al. 2002; Fruchter et al. 2006). GRBs are
located in host galaxies which are generally small, faint, blue and highly
star-forming (e.g. Chary et al. 2002; Fruchter et al. 1999; Le Floc'h et
al. 2003). Hence, detailed and extensive host galaxy observations provide a
wealth of information on the gas and dust properties of star-forming galaxies throughout cosmological history. 

Accurately measuring the dust content of these galaxies is of great importance
in, to name one example, the determination of their unobscured star formation
rates where uncertainty in the correction for dust can easily dominate the
errors on the measured star formation rates for high redshift galaxies (e.g. Pettini et al. 1998; Meurer et al. 1999).
Absorption within our own Galaxy along a particular line of sight can be
estimated and removed, but absorption which is intrinsic to the GRB host galaxy as a function of wavelength is unknown, and
is especially difficult to determine given its dependence on metallicity and
the possible existence of dusty intervening systems whose extinction
curves cannot be disentagled from those of the host galaxy. Afterglow
spectroscopy and/or photometry can be used to provide an estimate of the total
extinction along the line-of-sight to the GRB. If the host galaxy itself is
bright and extended enough to be observed once the afterglow has faded,
different lines-of-sight may be probed besides the off-centre UV-bright
regions within which GRBs are generally situated.

Extinction in the optical/UV
regime due to dust grains is typically modelled using either Milky Way (MW or Galactic),
Large Magellanic Cloud (LMC) or Small Magellanic Cloud (SMC) extinction curves
(e.g. Pei 1992) because these curves can be measured and so are well known, or with the Calzetti extinction law derived empirically from
UV observations of
starburst and blue compact galaxies (Calzetti et al. 1994). It appears that dust content in most GRB hosts produces an
SMC-like extinction law (e.g. Galama \& Wijers 2001; Vreeswijk et al. 2004; Stratta
et al. 2004; Kann et al. 2006; Schady et al. in preparation), owing to an observed lack
of the 2175\AA\ feature thought to be caused by carbonaceous dust
grains (Draine \& Lee 1984). This feature has, however, been clearly observed in
GRB afterglow spectra where the line of sight between us and the GRB is
intercepted by intervening systems: the best example to date, in which the extinction curve of an
intervening system could for the first time be disentangled from that of the
host galaxy, is GRB\,060418 (Ellison et al. 2006).

In general, low amounts of optical extinction are found towards GRBs,
unexpected if GRBs are located in dusty star-forming regions, whilst the X-ray
spectra reveal a different picture (first noted by Galama \& Wijers 2001). At
X-ray wavelengths absorption is caused by metals in both gas and solid phase,
predominantly oxygen and carbon (see e.g. Wilms et al. 2000), and
we often measure high values for the absorbtion columns. However, these absorption edges
are shifted out of the X-ray observing window for high redshifts ($z\gtrsim 2$)
beyond which only large columns can be measured and there is a degeneracy
between redshift and X-ray column density (e.g. Watson et al. 2002). 
The GRB host metallicity is observed to be
low (compared with the Milky Way) in measurements via optical spectroscopy of
a dozen or so afterglows. Host metallicities can reach values as low as 1/100
Solar (GRB\,050730, Starling et al. 2005; Chen et al. 2005, but see also
Prochaska 2006 for potential caveats) - even lower than found for the SMC (see Figure 3 of Fynbo et
al. 2006 for an overview). This only increases any measured X-ray column,
which is expressed as an equivalent hydrogen column density, $N_{\rm H}$. But
here we note that metallicities are not generally obtainable for lower
redshift GRBs ($z\lesssim 2$) due to the hydrogen Lyman-$\alpha$ line lying in
the far UV outside typical observing windows.

The apparent discrepancy between optical and X-ray extinction resulting in
high gas-to-dust ratios in GRB host galaxies (often far higher than for the
MW, LMC or SMC, e.g. GRB\,020124, Hjorth et al. 2003, but see Schady et
al. in preparation) is not satisfactorily explained, though the suggestion that dust
destruction can occur via the high energy radiation of the GRB (e.g. Waxman \&
Draine 2000) could possibly account for the discrepancy. 
It is thought that circumburst dust may be destroyed by sublimation of dust grains due to UV emission (Waxman \& Draine 2000), sputtering (Draine \& Salpeter 1979) or dust grain heating and charging (Fruchter et al. 2001).
Alternative models for the extinction by dust grains, including skewing the
dust grain size distribution towards larger grains, have been investigated,
and in fact such a grain size distribution may result from exposure of the
dust to the GRB radiation field since destruction of small grains is more
efficient than for larger grains (Perna et al. 2003). Attempts to
model the process of dust destruction have been made by e.g. Perna \& Lazzatti
(2002). However, such models have not replaced SMC-like (low metallicity)
extinction as the best description of most GRB environments (e.g. Stratta et
al. 2004; Kann et al. 2006).

Traditionally the optical and X-ray spectra have been treated seperately in
extinction studies. Since the underlying spectrum is likely a synchrotron
spectrum (power law (pl) or broken power law (bknpl), e.g. Galama \& Wijers 1998) extending through both wavelength regions, it would be most accurate
to perform simultaneous fits. More recently such fits have been made, either
by fitting the X-ray spectrum individually and thereby transforming the model counts
to flux to create a spectral energy distribution (SED) with the optical data in flux space (e.g. Stratta et
al. 2005; Watson et
al. 2006), or by using the {\it Swift} XRT X-ray and UVOT Ultraviolet (UV) and $U$,$B$,$V$ band data
together in a fit to the count spectrum (e.g. Blustin et al. 2005; Schady et
al. in preparation). The {\it Swift} UVOT data can be loaded directly into the xspec
spectral fitting package and treated in count space owing to its calibration,
which is generally not true for ground-based data. We
present here an alternative method, which makes use of
simultaneous fits in count space extending from near-infrared (nIR, in this
case $K$ band) to X-ray (10 keV) to
obtain the most accurate possible measurements of both the underlying
continuum spectrum and the extinction.  

In Section 4 we fit the broad-band SEDs (from nIR
through X-ray) of a subsample of the {\it BeppoSAX} sample of GRB afterglows
to better measure the extinction properties of their host galaxies - a sample
chosen for its availability of suitable data well studied in the
seperate optical and X-ray band passes.
Section 2 outlines the data sample and reduction techniques. Section 3
describes the method we use to model the broad-band SEDs and Section 4 presents
the results of our fitting through discussion of individual bursts and
comparison with previous studies. In Section 5 we discuss the implications of
our findings for galaxy extinction curves, and summarise and compare the
various methods now available to measure extinction in the hosts. We conclude
by summarising our method and findings in Section 6. An analysis of the blastwave parameters and density profiles for the circumburst media obtained from these fits will be presented in a forthcoming paper (Starling et al., Paper II).

\section{Observations}
This sample of 10 long GRBs observed with the {\it BeppoSAX} Narrow Field
Instruments (approx. 0.1--10 keV) is chosen for the good
availability (3 bands or more) of optical/nIR photometry (Tables 1 and 2). The
optical/nIR bands available for each source and their references are listed in
Table~\ref{tbl-3}. As these
GRBs are all previously studied, overlapping with the
samples studied by Galama \& Wijers (2001) and Stratta et al. (2004), this constitutes a good sample on which to
first adopt this method of simultaneous SED fitting.

All X-ray observations are taken from the {\it BeppoSAX} data archives, using
the LECS and MECS instruments raw data within the energy ranges 0.1--4 and
1.0--10 keV respectively. Data have been reduced using the {\small SAXDAS}
routines. We combined data from the MECS2 and MECS3 instruments (except in the
case of GRB\,970228, where we use the MECS3 instrument only, see Stratta et
al. 2004), including a gain
equalisation. We then combined multiple observations for each source and
instrument type among the narrow field instruments, omitting the last
observation if it was $\ge$ 3 days later than the previous one, before
extracting spectra.
Background X-ray spectra were taken from blank fields and count rates
checked against the local background finding no adjustments necessary: the net
count rates of the two types of field agree on average to within 0.0001 counts
s$^{-1}$. The latest canned arf and rmf files were used with MECS data, but
were created for LECS observations at their off-axis source positions (listed
in Table 3 of Stratta et al. 2004).
We group all spectra such that a minimum of 20 counts are in each bin in order to use the $\chi^2$ statistic. However, in the case of GRB\,000926 there were very few counts in the X-ray spectrum so we have required only 10 counts per bin. 
There is a known offset between the normalisations of the LECS and MECS
instruments. We fit for this offset in the X-ray spectra only, adding a
constant-value free parameter to the model and adopting MECS as the reference
for the LECS instrument. We fix the offset values in the SED of each GRB to these values (Table~\ref{tbl-1}).

All temporal decay slopes, both for X-ray and optical lightcurves, have been taken from the literature and are listed together
with Galactic extinction corrections in Table~\ref{tbl-2}. Optical and nIR photometry
was taken from the literature and from our own nIR observations of GRB\,990510
(described in Curran et al. in preparation).

\begin{table*}
\begin{center}
\caption{\emph{BeppoSAX} data table for combined datasets. We
  have measured the MECS--LECS offset values from the combined X-ray spectra
  with respect to the MECS instrument.
\label{tbl-1}}
\begin{tabular}{c c c c c c c c}
\hline \hline
GRB & obs start & obs end & obs midpt(log) & \multicolumn{2}{c}{$t_{\rm on-source}$ (s)} &MECS--LECS offset\\
    &    \multicolumn{3}{c}{(days since trigger)}    &     MECS & LECS    &\\
\hline \hline
970228 & 0.344& 0.693&0.520&1.43$\times$10$^{4}$& 5.611$\times$10$^{3}$&1.45$^{+0.67}_{-0.42}$\\
970508 & 0.434& 3.091&1.679&5.90$\times$10$^{4}$&2.36$\times$10$^{4}$&1.09$^{+0.50}_{-0.43}$\\
971214 & 0.274& 2.528& 1.362& 1.01$\times$10$^{5}$ & 4.62$\times$10$^{4}$& 1.57$^{+0.67}_{-0.56}$\\
980329 & 0.294& 2.026 & 1.148 & 6.38$\times$10$^{4}$ & 2.49$\times$10$^{4}$ &0.68$^{+0.58}_{-0.36}$\\
980519 & 0.406 &1.468&0.930& 7.82$\times$10$^{4}$&2.31$\times$10$^{4}$&1.00$^{+0.94}_{-0.64}$\\
980703 & 0.827& 1.902& 1.333& 3.92$\times$10$^{4}$& 1.66$\times$10$^{4}$&0.99$^{+0.36}_{-0.34}$ \\ 
990123 & 0.242 &  2.573 &  1.245 &8.20$\times$10$^{4}$ & 2.80$\times$10$^{4}$ & 0.77$\pm$0.9   \\
990510 & 0.334 &  1.850 & 1.067 & 6.79$\times$10$^{4}$ & 3.17$\times$10$^{4}$ & 0.86$^{+0.13}_{-0.12}$\\
000926 & 2.003 &2.466& 2.234 &  1.96$\times$10$^{4}$ &5.027$\times$10$^{3}$&0.70$^{+2.13}_{-0.67}$\\
010222 & 0.376 & 2.703 & 1.511& 8.84$\times$10$^{4}$& 5.11$\times$10$^{4}$ &1.43$^{+0.14}_{-0.13}$\\
\hline
\end{tabular}  
\end{center}
\end{table*}

\begin{sidewaystable*}
\caption{GRB known properties: Galactic absorption (columns 2-3, $\times$10$^{22}$ cm$^{-2}$, and column 4), redshift,
  optical temporal decay slope(s) and (jet) break time in days since trigger and X-ray temporal slope.}
\label{tbl-2}
\begin{center}
\begin{tabular}{c c c c c c c c c}

\hline\hline
GRB & Gal. $N_{\rm H}$$^{(1)}$ & Gal. $N_{\rm H}$$^{(2)}$&$E(B-V)_{\rm Gal}$& redshift $z$ &$\alpha_{1}$$^{*}$&$\alpha_{2}$$^{*}$&$t_{\rm bk}$$^{*}$&$\alpha_x$$^{+}$\\
\hline\hline
970228 &0.165& 0.134& 0.203&0.6950$\pm$0.0003$^{\mathrm{a}}$& 1.46$\pm$0.15&-&-&1.3$\pm$0.2\\
970508 &0.0526&0.0485& 0.050&0.835$\pm$0.001$^{\mathrm{b}}$&1.24$\pm$0.01&-&-&1.1$\pm$0.1\\
971214 &0.0167&0.0128& 0.016&3.418$\pm$0.010$^{\mathrm{c}}$&1.49$\pm$0.08&-&-&1.6$\pm$0.1\\
980329 &0.0918&0.0916& 0.073&3.6$^{\mathrm{d}}$ &0.85$\pm$0.12&-&-&1.5$\pm$0.2\\
980519 &0.183&0.189&0.267 &- &1.50$\pm$0.12&2.27$\pm$0.03&0.48$\pm$0.03&1.83$\pm$0.3\\
980703 &0.0579&0.0498&0.057 &0.9661$\pm$0.0001$^{\mathrm{e}}$&0.85$\pm$0.84&1.65$\pm$0.46&1.35$\pm$0.94&0.9$\pm$0.2 \\
990123 &0.0213&0.0165&0.016 &1.600$\pm$0.001$^{\mathrm{f}}$& 1.24$\pm$0.06&1.62$\pm$0.15&2.06$\pm$0.83&1.44$\pm$0.11\\
990510 &0.0924&0.0815&0.203&1.619$\pm$0.002$^{\mathrm{g}}$&0.92$\pm$0.02 &2.10$\pm$0.06& 1.31$\pm$0.07&1.4$\pm$0.1\\
000926 &0.0265&0.0220&0.023&2.0379$\pm$0.0008$^{\mathrm{h}}$& 1.74$\pm$0.03 & 2.45$\pm$0.05 & 2.10$\pm$0.15&1.7$\pm$0.5\\
010222 &0.0163&0.0175&0.023&1.4768$\pm$0.0002$^{\mathrm{i}}$&0.60$\pm$0.09&1.44$\pm$0.02&0.64$\pm$0.09 & 1.33$\pm$0.04\\
\hline
\end{tabular} 
\end{center}
 \begin{list}{}{} 
    \item[$^{(1)}$]taken from Dickey \& Lockman 1990 (resolution of $\sim$1$^{\circ}$)
    \item[$^{(2)}$]taken from the Leiden/Argentine/Bonn Galactic H~I Survey, Kalberla et al. 2005 (resolution of $\sim$0.6$^{\circ}$)
    \item[$^{*}$]taken from Zeh et al. 2006, where uncertainties are 1$\sigma$
    \item[$^{+}$]taken from Gendre \& Boer 2005 and in 't Zand et al. 1998
      (GRB\,980329, MECS 2--10 keV data), Nicastro et al. 1999 (GRB\,980519)
    \item[$^{\mathrm{a}}$]Bloom et al. 2001~~$^{\mathrm{b}}$~Bloom et
      al. 1998a~~$^{\mathrm{c}}$~Kulkarni et al. 1998~~$^{\mathrm{d}}$~photometric redshift only, Jaunsen et al. 2003~~$^{\mathrm{e}}$~Djorgovski et al. 1998~~$^{\mathrm{f}}$~Kulkarni et al. 1999a~~$^{\mathrm{g}}$~Vreeswijk et al. 2001~~$^{\mathrm{h}}$~Castro et al. 2003~~$^{\mathrm{i}}$~Mirabal et al. 2002
  \end{list}
\end{sidewaystable*}

\section{Method}
Per source all data are fitted simultaneously, assuming wherever possible no
prior model. This is achieved by fitting in count space (as is traditional in
the X-ray regime where one fits for the emission model, extinction and instrumental response simultaneously): the optical and nIR magnitudes are converted to flux and
then to counts. For the magnitude to flux conversion
we use the zero points and effective bandwidths of each optical band
(Johnson for $U$, $B$, $V$, $R$, $I$, $J$, $K$, 2MASS for $H$ and $K_s$ and
Bessel for $V_c$, $R_c$, $I_c$). In the small number of cases for which
the specific band is not stated, we assume the appropriate Johnson filter.
These fluxes are then converted to photons cm$^{-2}$ s$^{-1}$ per bin (bin
width $=$ effective bandwidth of the filter) within
the {\small ISIS} spectral fitting program (Houck \& Denicola 2000) which is
equivalent to the X-ray units of counts cm$^{-2}$ s$^{-1}$ per
bin since total number of counts is conserved.

Since we fit in count space, we need not first assume a model for the X-ray spectrum to
convert the counts to flux. Herein lies one advantage of using this
method. The second advantage comes through the multiwavelength approach. The
nIR, optical and X-ray spectra are related since we assume the broadband
spectrum is caused by synchrotron emission, hence a simultaneous fit provides
greater accuracy and consistency between the parameters. Inclusion of nIR data
and $R$ band optical data together with the 2--10 keV X-ray data, regions over
which
extinction has the least effect, allows the underlying power law slope to be most
accurately determined.

The X-ray data typically have much longer exposure times than the individual
optical/IR measurements, particularly given that where sensible we have
combined X-ray datasets from different epochs to increase the signal to noise. We have chosen to
omit the last X-ray observation of any GRB which occurs more than 3 days after
the previous observation where this contributes more to the
background noise than the signal and skews the observation time midpoint. We
note that structure in the X-ray lightcurves and spectral changes at early
times have often been reported for the better and earlier sampled {\it Swift}
GRB afterglows (e.g. Nousek et al. 2006), which may affect the {\it BeppoSAX} afterglows with fairly early coverage, i.e. those beginning at 0.2--0.3 days since burst (see Table~\ref{tbl-1}).
All optical data are scaled to a common time, which
corresponds to the midpoint of the X-ray observations, calculated in
log-space. This is done by extrapolation according to the decay rates and
lightcurve break times determined in a thorough analysis of pre-{\it Swift}
burst optical lightcurves by Zeh et al. (2006, Table~\ref{tbl-2}). 

We fit for the X-ray and optical extinction at the redshift of the GRB. For GRBs 980329 and 980519 the redshift is not known,
and therefore absorption at the source cannot be measured. However, a
photometric redshift has been made for GRB\,980329 of $z\sim$ 3.6 (Jaunsen et
al. 2003) which we adopt here. For GRB\,980519 we adopt the mean redshift for
this sample, $z\sim$ 1.58, to make an estimate of the intrinsic absorption
required. 

Flux is depleted in the bluemost optical bands for high redshift bursts due to
the Lyman absorption edges and the resulting transmission is calculated for
each band per burst (see Curran et al. in preparation for further details of
this calculation, which we note involves assuming a spectral slope for the
optical flux over a given band and uses parameters from Madau 1995 and Madau
et al. 1996). However, this has an effect only on the $U$ through $R$ band
magnitudes of GRBs 971214 and 980329 and a minor effect on the $U$ band magnitudes of GRBs 980519, 990123, 990510 and 000926.

The errors on the optical magnitudes are taken from the photometric errors in
the literature, and are set to 0.1 for cases where the literature reports a
smaller error to account for systematic uncertainties. The conversion from
magnitudes to flux has an associated error of up to 5
\% (Fukugita et al. 1995). We note that by
extrapolating magnitudes to different times we introduce a possible random
error, since there is an uncertainty in the measured decay indices and break
times, which we can allow for in an offset between the optical and X-ray
data. This error is not included in the individual data points because this
would introduce an artificially large error in the optical slope which is not
in fact present since the relative errors between the bands do not change.

\begin{table}
\caption{Range of optical and nIR magnitudes used in this study.}
\label{tbl-3}
\begin{center}  
\begin{tabular}{c c c}
\hline\hline
GRB & bands used & refs\\
\hline\hline
970228 & \emph{VR$_{c}$I$_{c}$} & [1]\\
970508 & \emph{UBVR$_{c}$I$_{c}$K$_{s}$} & [2,3]\\
971214 & \emph{VRIJK$_{s}$} & [4,5,6,7]\\
980329 & \emph{RIJK} & [8]\\
980519 & \emph{UBVR$_{c}$I$_{c}$} &[9]\\
980703 & \emph{RIJHK} &[10]\\
990123 & \emph{UBVRIHK}& [11]\\
990510 & \emph{BVRIJHK$_{s}$} &[12,13]\\
000926 & \emph{UBVRIJHK} &[14]\\
010222 & \emph{UBVRIJ} &[15]\\
\hline
\end{tabular} 
\end{center}
  \begin{list}{}{} 
    \item[1]Galama et al. 2000~2~Galama et al. 1998b~3~Chary et al. 1998~4~Halpern et al. 1998~5~Diercks et al. 1997~6~Tanvir et al. 1997~7~Ramaprakash et al. 1998~8~Reichart  et al. 1999 and references therein (their Table 1)~9~Jaunsen et al. 2001~10~Vreeswijk et al. 1999~11~Galama et al. 1999~12~Stanek et al. 1999~13~Curran et al. in prep.~14~Fynbo et al. 2001~15~Masetti et al. 2001
  \end{list}
\end{table}

\subsection{Models}
\begin{figure}
\begin{center}
\includegraphics[width=6cm, angle=90]{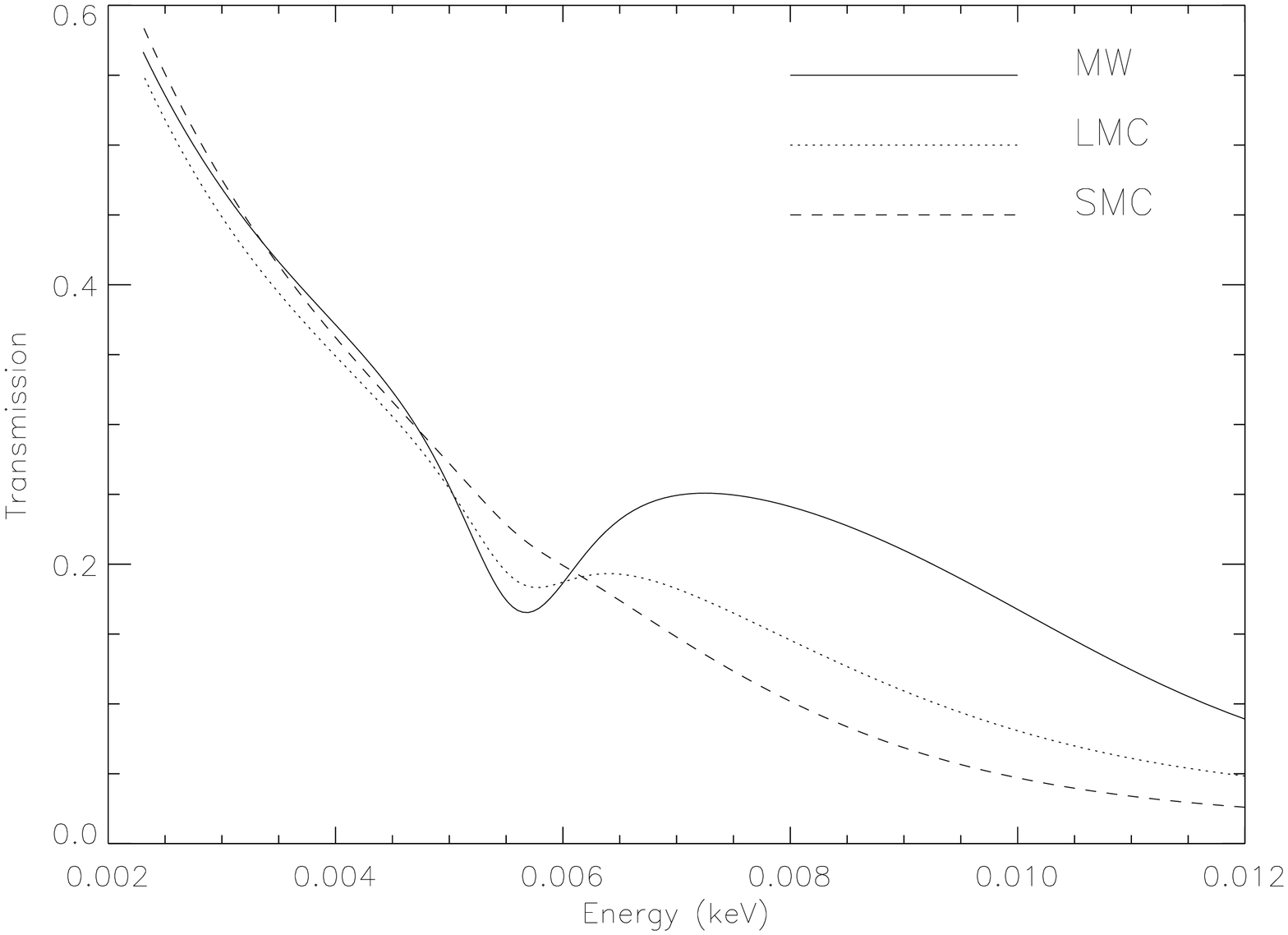}
\caption{The well known extinction curves for the Milky Way (MW), Large and
  Small Magellanic Clouds (LMC and SMC respectively, Pei 1992). The transmission of
  optical/UV intrinsic flux with energy is shown for an object at redshift
  $z=0$ and with a large optical extinction of $E(B-V)=0.2$.
\label{fig1}}
\end{center}
\end{figure}

Data are fitted within the spectral fitting package {\small ISIS} (Houck \& Denicola 2000) using both models written for use within xspec (Arnaud et al. 1996) and models written for {\small ISIS}. 
Models consist of either a single or a broken power law, to allow for a
possible cooling break in between the optical and X-rays. Should the break in
the power law be due to cooling, the difference in slope is $\Delta\beta=0.5$
(e.g. Wijers \& Galama 1999), which we fix in the broken power law model. \\
On its way to us, the intrinsic power law is absorbed by optical extinction
at the host/burst redshift (this could be local to the GRB itself or another
location within the host galaxy). The extinction curves used for intrinsic
optical extinction in this study are Galactic-, SMC- and LMC-like (Figure~\ref{fig1})
following the prescriptions of Cardelli et al. (1989) and Pei (1992)
respectively. We do not use the Calzetti extinction curve (Calzetti et
al. 1994) because it has a
larger error associated with it, being constructed from fewer measurements
than those for the nearby Magellanic Clouds. There is also absorption in the
X-ray regime predominantly by metals, e.g. the oxygen edge, for which we use a
photoelectric absorption model. We refer the reader to Wilms et al.
(2000) for a detailed description of the X-ray absorption models. The X-ray
absorption model can be computed for various metallicities by simple scaling
of the Solar abundances by a constant factor. We adopt firstly Solar
metallicity and secondly the metallicity assumed in the optical extinction
model: using SMC-like absorption one would adopt $Z$=1/8 Z$_\odot$ and for
LMC-like absorption $Z$=1/3 Z$_\odot$ (Pei 1992), for self-consistency.

The flux is then corrected for Galactic absorption (Table~\ref{tbl-1}). In the X-ray
regime these values are fixed at the $N_{\rm H}$ values given in Dickey \&
Lockman (1990), which are averages over 1 degree at the positions as given in
the Simbad catalogues or from the {\it BeppoSAX} Narrow-field instruments. For
completeness and comparison, we also list in Table~\ref{tbl-1} extinction measurements
from the newer and slightly higher resolution Leiden/Argentine/Bonn Galactic
H~I Survey (resolution of $\sim$0.6$^{\circ}$, Kalberla et al. 2005). These
new values are not significantly different than those of Dickey \& Lockman
though appear to be generally lower, and
to date all previous studies use the values from Dickey \& Lockman which we will also use here. For
the optical extinction we use $E(B-V)_{\rm Gal}$ values given by Schlegel et al. (1998) from their full-sky 100 \micron\ map together with
the Galactic extinction curve of Cardelli et al. (1989) with
$R_V=A_V/E(B-V)=3.1$. The Schlegel et al. (1998) maps have a best resolution of 6.1': for each source we use the best resolution available, but in some cases we must use an average over 1 square degree centred on the source coordinates.\\
The fit statistic calculated is $\chi^2$, using a Levenberg-Marquardt fit
minimisation method. Errors in the LECS-MECS offsets (except for
GRB\,970508), optical decay slopes and redshifts are not propagated through
the fitting routines. These values are simply fixed from Tables 1 and 2
respectively. We also do not include uncertainties on the zero points on our
photometric data points (see previous Section).

\begin{sidewaystable*}
\caption{~Results (main parameters) of fits to the spectral energy
  distributions. For each GRB we fitted the SED with the power law and the broken
  power law models with all 3 extinction models (MW, LMC and SMC). X-ray column densities given in brackets are for the appropriate metallicities (LMC or SMC), otherwise Solar metallicity is assumed. The LECS-MECS offsets were fixed at the central values listed in Table 2. For the broken power law models $\Gamma_1$ = $\Gamma_2$ - 0.5. Where the break energies are unconstrained (indicated by the letter U) we give in brackets the central value derived and any limits set. All errors are quoted at the 90 \% confidence level (or 1.6$\sigma$). The F-test probability gives the probability that the result is obtained by chance, therefore a significant improvement in the fit when adding one extra free parameter is indicated by a low probability. For GRB\,970508, where two F-test probabilities are listed the second is for the comparison between fits with the optical--X-ray offset free and fixed. Plots of the data overlaid with their best-fitting models are shown in Figure 2.}
\label{tbl-4}
\renewcommand{\arraystretch}{1.2}
\begin{center}
\begin{tabular}{l l l l l l l}
\hline\hline
GRB/model & $N_{\rm H, int}$ ($\times$10$^{22}$ atoms cm$^{-2}$)& $E(B-V)_{\rm int}$& $\Gamma_{(2)}$ & $E_{\rm bk}$ (keV)&$\chi^2$/dof& F-test prob. \\ \hline
GRB\,970228 & & & & & &\\
\hline
pl+mw&0$^{+0.58}_{-0}$&0$^{+0.20}_{-0}$&1.72$^{+0.11}_{-0.03}$& &11.6/14&\\
pl+lmc&0$^{+0.57}_{-0}$ (0.001$^{+1.40}_{-0.001}$)&0$^{+0.17}_{-0}$&1.72$^{+0.10}_{-0.03}$& &11.6/14&\\
pl+smc&0$^{+0.57}_{-0}$ (0.01$^{+2.48}_{-0.01}$)&0$^{+0.17}_{-0}$&1.72$^{+0.09}_{-0.03}$& &11.6/14&\\
bknpl+mw&0.54$^{+0.68}_{-0.46}$&0$^{+0.19}_{-0}$&2.06$^{+0}_{-0.03}$&0.32$^{+0.09}_{-0.32}$ &7.6/13&2.1$\times$10$^{-2}$\\ 
bknpl+lmc&0.54$^{+0.68}_{-0.41}$ (1.22$^{+1.65}_{-0.97}$)&0$^{+0.16}_{-0}$&2.06$^{+0.23}_{-0.03}$&0.32$^{+0.09}_{-0.32}$ &7.6/13&2.1$\times$10$^{-2}$\\ 
bknpl+smc&0.54$^{+0.68}_{-0.41}$ (2.01$^{+2.96}_{-1.66}$)&0$^{+0.16}_{-0}$&2.06$^{+0}_{-0.03}$&0.32$^{+0.09}_{-0.32}$ &7.6/13&2.1$\times$10$^{-2}$\\ 
\hline \hline
GRB\,970508 & & & & & &\\
\hline
pl+mw&1.57$^{+1.33}_{-0.85}$&0$^{+0.008}_{-0}$&1.94$^{+0.02}_{-0.01}$& &38.7/31& \\
pl+lmc&1.57$^{+1.33}_{-0.85}$ (3.83$^{+3.45}_{-2.14}$)&0$^{+0.007}_{-0}$&1.94$^{+0.02}_{-0.01}$&&38.7/31& \\
pl+smc&1.57$^{+1.33}_{-0.85}$ (6.84$^{+6.72}_{-4.01}$)&0$^{+0.007}_{-0}$&1.94$^{+0.02}_{-0.01}$&&38.7/31 &\\
bknpl+mw&1.93$^{+0.09}_{-0.90}$&0.043$^{+0.015}_{-0.043}$&2.09$^{+0.19}_{-0.01}$&U ($<$0.27)&31.2/30&1.2$\times$10$^{-2}$\\
bknpl+lmc& 1.93$^{+0.05}_{-0.90}$&0.040$^{+0.014}_{-0.040}$&2.09$^{+0.18}_{-0.01}$&U ($<$0.25)&31.7/30&1.5$\times$10$^{-2}$\\
bknpl+smc&2.08$^{+0.58}_{-1.28}$&0.035$\pm$0.035&2.12$^{+0}_{-0.06}$&U ($<$0.26) &32.2/30&2.1$\times$10$^{-2}$\\
\hline
optical:X-ray &offset&free&(see text)&&&\\
\hline 
pl+mw& 0.718$^{+1.39}_{-0.718}$& 0.004$^{+0.060}_{-0.004}$&1.76$^{+0.07}_{-0.24}$&&32.1/30&1.9$\times$10$^{-2}$\\
pl+lmc& 0.756$^{+1.34}_{-0.626}$ (1.80$^{+3.41}_{-1.52}$)&0$^{+0.054}_{-0}$& 1.78$^{+0.05}_{-0.23}$& &32.1/30&1.9$\times$10$^{-2}$\\
pl+smc&0.756$^{+1.07}_{-0.626}$ (3.50$^{+6.36}_{-2.89}$)&0$^{+0.052}_{-0}$ (0.005$^{+0.027}_{-0.005}$)&1.78$^{+0}_{-0.01}$&&32.1/30&1.9$\times$10$^{-2}$\\
bknpl+mw&0.63$^{+0.49}_{-0.63}$&0.032$^{+0.071}_{-0.032}$& 2.14$^{+0.14}_{-0.29}$&U (3.42)&29.7/29&0.236,0.137\\
bknpl+lmc&0.72$^{+0.05}_{-0.72}$&0.021$^{+0.022}_{-0.021}$&2.17$^{+0.14}_{-0.325}$&U (3.48)&30.1/29&0.224,0.176\\
bknpl+smc&0.88$^{+0.09}_{-0.88}$&0.007$^{+0.090}_{-0.007}$&2.23$^{+0.08}_{-0.34}$&U (3.56) &30.3/29&0.188,0.200\\
\end{tabular} 
\end{center}
\end{sidewaystable*}

\addtocounter{table}{-1}
\begin{sidewaystable*}
\caption{-- {\it continued}}
\renewcommand{\arraystretch}{1.2}
\begin{center}
\begin{tabular}{l l l l l l l}
GRB\,971214 & & & & & &\\
\hline
pl+mw&10.88$^{+15.97}_{-10.16}$ &0.045$^{+0.038}_{-0.042}$& 1.60$\pm$0.04&& 37.6/44& \\
pl+lmc&11.47$^{+16.16}_{-10.33}$ (31.70$^{+45.78}_{-28.42}$)&0.036$^{+0.021}_{-0.023}$& 1.60$\pm$0.04&& 34.4/44& \\
pl+smc&11.48$^{+16.15}_{-10.33}$ (68.77$^{+104.3}_{-61.26}$)&0.031$\pm$0.018&
1.60$\pm$0.03&& 33.2/44& \\
bknpl+mw& 28.37$^{+20.45}_{-26.53}$&0.056$\pm$0.051&2.04$^{+0.09}_{-0.14}$&U (1.56) &35.8/43 &0.149\\ 
bknpl+lmc& 29.78$^{+20.30}_{-25.94}$&0.044$^{+0.057}_{-0.027}$&2.04$^{+0.09}_{-0.14}$&U (1.40)& 31.7/43 &6.2$\times$10$^{-2}$\\ 
bknpl+smc&22.12$^{+26.61}_{-15.38}$&0.058$^{+0.027}_{-0.040}$&1.85$^{+0}_{-0.12}$& U (0.048)& 30.1/43&4.1$\times$10$^{-2}$ \\ 
\hline \hline
GRB\,980329 & & & & & &\\
\hline
pl+mw& 0.0012$^{+9.74}_{-0.0012}$&0.358$^{+0.088}_{-0.050}$&1.88$^{+0.07}_{-0.09}$& & 38.4/27 &\\
pl+mw+$z$ free&0.001$^{+9.52}_{-0.001}$&0.247$^{+0.039}_{-0.042}$&1.81$^{+0.04}_{-0.05}$& & 32.7/26&4.3$\times$10$^{-2}$\\ 
pl+lmc&0.001$^{+8.6}_{-0.001}$ (0.001$^{+24.2}_{-0.001}$)&0.210$^{+0.047}_{-0.019}$&1.84$^{+0.06}_{-0.07}$& & 35.4/27& \\
pl+smc& 0$^{+8.2}_{-0}$ (0.001$^{+51.3}_{-0.001}$)&0.178$^{+0.039}_{-0.030}$&1.82$^{+0.04}_{-0.07}$& & 34.3/27& \\
bknpl+mw&3.42$^{+21.9}_{-3.42}$&0.346$^{+0.148}_{-0.120}$&2.34$^{+0.08}_{-0.34}$&U (2.39: $<$6.89)&35.4/26&0.150 \\ 
bknpl+lmc& 8.00$^{+15.7}_{-8.00}$ (21.7$^{+43.9}_{-21.7}$)&0.211$^{+0.080}_{-0.064}$&2.26$\pm$0.31&U (1.09: $<$5.79)&31.7/26 &9.3$\times$10$^{-2}$\\
bknpl+smc& 7.50$^{+15.4}_{-7.50}$ (42.5$^{+95.9}_{-42.5}$)&0.179$^{+0.067}_{-0.051}$&2.25$^{+0.10}_{-0.31}$&U (1.08: $<$5.54)&30.2/26&7.2$\times$10$^{-2}$ \\
\hline \hline
GRB\,980519 & & & & & &\\
\hline
pl+mw&0$^{+5.4}_{-0}$&0.008$^{+0.015}_{-0.008}$&1.97$^{+0.05}_{-0.03}$& & 18.9/23&\\ 
pl+lmc&0$^{+5.8}_{-0}$ (0.06$^{+15.12}_{-0.06}$)&0.014$^{+0.024}_{-0.004}$&1.98$\pm$0.04& & 18.0/23&\\
pl+smc&0.005$^{+5.9}_{-0.005}$ (0.36$^{+29.6}_{-0.36}$)&0.013$^{+0.019}_{-0.009}$&1.98$^{+0.03}_{-0.04}$& & 17.7/23&\\
bknpl+mw&0.84$^{+1.96}_{-0.84}$&0.012$^{+0.045}_{-0.012}$&2.44$^{+0.07}_{-0.10}$&U (1.76: $>$0.26)&17.4/22&0.182\\ 
bknpl+lmc&1.27$^{+0.86}_{-1.27}$ (3.22$^{+21.9}_{-3.22}$)&0.019$^{+0.055}_{-0.019}$&2.44$^{+0.07}_{-0.10}$&U (1.56)&16.0/22&0.111\\
bknpl+smc&1.39$^{+0.63}_{-1.39}$ (6.52$^{+43.0}_{-6.52}$)&0.017$^{+0.038}_{-0.017}$&2.43$^{+0.07}_{-0.10}$&U (1.46)&15.6/22&9.9$\times$10$^{-2}$\\
\hline \hline
GRB\,980703 & & & & & & \\
\hline
pl+mw&0.55$^{+1.02}_{-0.55}$&0.302$\pm$0.059 &1.92$\pm$0.03 && 30.2/27 & \\
pl+lmc&0.54$^{+1.02}_{-0.54}$ (1.33$^{+2.53}_{-1.33}$)&0.275$\pm$0.054 &1.92$\pm$0.03 && 30.0/27 &\\
pl+smc&0.53$^{+1.01}_{-0.53}$ (2.35$^{+4.64}_{-2.35}$)&0.287$^{+0.057}_{-0.056}$ &1.92$\pm$0.03 && 29.8/27 &\\
bknpl+mw&1.35$^{+1.47}_{-1.06}$&0.31$^{+0.09}_{-0.06}$&2.38$^{+0.06}_{-0.24}$&1.40$^{+1.84}_{-1.38}$& 22.9/26 &8$\times$10$^{-3}$\\ 
bknpl+lmc&1.34$^{+1.47}_{-1.06}$ (3.28$^{+3.70}_{-2.60}$)&0.28$^{+0.08}_{-0.06}$&2.37$^{+0.05}_{-0.24}$&1.40$^{+1.83}_{-1.38}$& 22.6/26&7$\times$10$^{-3}$ \\ 
bknpl+smc&1.33$^{+1.46}_{-1.05}$ (5.80$^{+6.87}_{-4.64}$)&0.30$^{+0.08}_{-0.06}$&2.37$^{+0.05}_{-0.24}$&1.40$^{+1.81}_{-1.38}$& 22.3/26&7$\times$10$^{-3}$ \\ 
\end{tabular} 
\end{center}
\end{sidewaystable*}

\addtocounter{table}{-1}
\begin{sidewaystable*}
\caption{-- {\it continued}}
\renewcommand{\arraystretch}{1.2}
\begin{center}
\begin{tabular}{l l l l l l l}
GRB\,990123 & & & & & &\\
\hline
pl+mw&0$^{+0.11}_{-0}$&0.006$^{+0.019}_{-0.002}$&1.61$\pm$0.01&& 191/121 &\\ 
pl+lmc&0$^{+0.11}_{-0}$&0.004$^{+0.013}_{-0.004}$&1.61$\pm$0.01&& 191/121& \\
pl+smc& 0$^{+0.11}_{-0}$&0.004$^{+0.018}_{-0.004}$&1.61$\pm$0.01&& 191.3/121 &\\
bknpl+mw& 0.61$^{+0.51}_{-0.49}$&0.01$^{+0.03}_{-0.01}$&2.01$^{+0}_{-0.04}$&0.67$^{+1.74}_{-0.49}$& 115.5/120 &7$\times$10$^{-15}$\\
bknpl+lmc&0.59$^{+0.52}_{-0.37}$(1.43$^{+1.33}_{-1.42}$)&0.01$^{+0.02}_{-0.01}$&2.00$^{+0}_{-0.04}$&0.55$^{+1.85}_{-0.40}$&115/120 &7$\times$10$^{-15}$\\
bknpl+smc&0.59$^{+0.51}_{-0.37}$(2.54$^{+2.43}_{-2.54}$)&0.014$^{+0.024}_{-0.014}$&1.99$^{+0}_{-0.04}$&0.55$^{+1.85}_{-0.40}$&115/120 &6$\times$10$^{-15}$\\
\hline \hline
GRB\,990510 & & & & & &\\
\hline
pl+mw& 0$^{+0.286}_{-0}$& 0$^{+0.003}_{-0}$ &1.855$^{+0.010}_{-0.007}$& &129/78 &\\
pl+lmc&0$^{+0.344}_{-0}$& 0$^{+0.003}_{-0}$&1.854$^{+0.009}_{-0.010}$&& 129/78 &\\
pl+smc&0$^{+0.340}_{-0}$& 0$^{+0.003}_{-0}$&1.855$^{+0.007}_{-0.010}$&&129/78 &\\
bknpl+mw&0.12$^{+0.75}_{-0.12}$&0$^{+0.01}_{-0}$& 2.03$^{+0}_{-0.01}$&0.018$\pm$0.002& 83.3/77 &7$\times$10$^{-9}$\\ 
bknpl+lmc&0.13$^{+0.02}_{-0.13}$ (0.34$^{+1.88}_{-0.34}$)&0$^{+0.01}_{-0}$& 2.03$^{+0.07}_{-0.01}$&0.018$\pm$0.002& 83.3/77&7$\times$10$^{-9}$ \\ 
bknpl+smc&0.13$^{+0.02}_{-0.13}$ (0.74$^{+3.26}_{-0.74}$)&0$^{+0.01}_{-0}$& 2.03$^{+0.07}_{-0.01}$&0.018$\pm$0.002& 83.3/77&7$\times$10$^{-9}$ \\ 
\hline \hline
GRB\,000926 & & & & & &\\
\hline
pl+mw & 0 fixed& 0.166$^{+0.021}_{-0.025}$&1.80$^{+0.09}_{-0.05}$& & 27.5/16 &\\
pl+lmc& 0 fixed& 0.119$\pm$0.015&1.77$^{+0.07}_{-0.04}$& & 11.7/16 &\\
pl+smc& 0 fixed& 0.100$^{+0.014}_{-0.015}$&1.76$^{+0.08}_{-0.05}$& & 15.0/16& \\
bknpl+mw& 0 fixed&0.167$^{+0.026}_{-0.025}$&2.30$\pm$0.09&U (4.36: $>$0.59)& 27.4/15&0.818 \\
bknpl+lmc&0 fixed&0.122$^{+0.044}_{-0.017}$&2.25$^{+0.08}_{-0.28}$&U (2.23: $>$0.02)& 11.1/15&0.382 \\
bknpl+smc&0 fixed&0.102$^{+0.017}_{-0.016}$&2.25$^{+0.09}_{-0.11}$&U (2.90: $>$0.38)& 14.7/15&0.588 \\
\hline \hline
GRB\,010222 & & & & & &\\
\hline
pl+mw& 0.63$^{+0.30}_{-0.24}$&0.063$\pm$0.033&1.87$\pm$0.03&&  97.2/137& \\
pl+lmc&0.60$^{+0.29}_{-0.23}$ (1.44$^{+0.72}_{-0.566}$)& 0.043$^{+0.022}_{-0.021}$& 1.86$^{+0.024}_{-0.025}$&& 95.5/137& \\
pl+smc&0.58$^{+0.28}_{-0.22}$ (2.39$^{+1.28}_{-0.987}$)& 0.035$^{+0.019}_{-0.018}$ & 1.85$^{+0.021}_{-0.022}$ && 96.3/137 &\\
bknpl+mw& 1.35$^{+0.57}_{-0.47}$& 0.087$\pm$0.051& 2.07$^{+0.09}_{-0.08}$& U (0.03)&96.9/136& 0.518\\
bknpl+lmc& 1.15$^{+0.53}_{-0.36}$ (2.68$^{+1.33}_{-0.813}$) & 0.076$^{+0.025}_{-0.029}$& 2.02$^{+0.09}_{-0.04}$ & 0.01$^{+0.05}_{-0.01}$& 86.5/136 &2$\times$10$^{-4}$\\ 
bknpl+smc& 1.15$^{+0.54}_{-0.40}$ (4.57$^{+2.37}_{-1.54}$) & 0.060$\pm$0.023 & 2.02$^{+0.09}_{-0.05}$ & 0.017$^{+0.063}_{-0.017}$ &86.3/136 &1$\times$10$^{-4}$\\ 
\hline
\end{tabular} 
\end{center}
\end{sidewaystable*}

\begin{figure*}
\begin{center}
\includegraphics[width=6cm, angle=0]{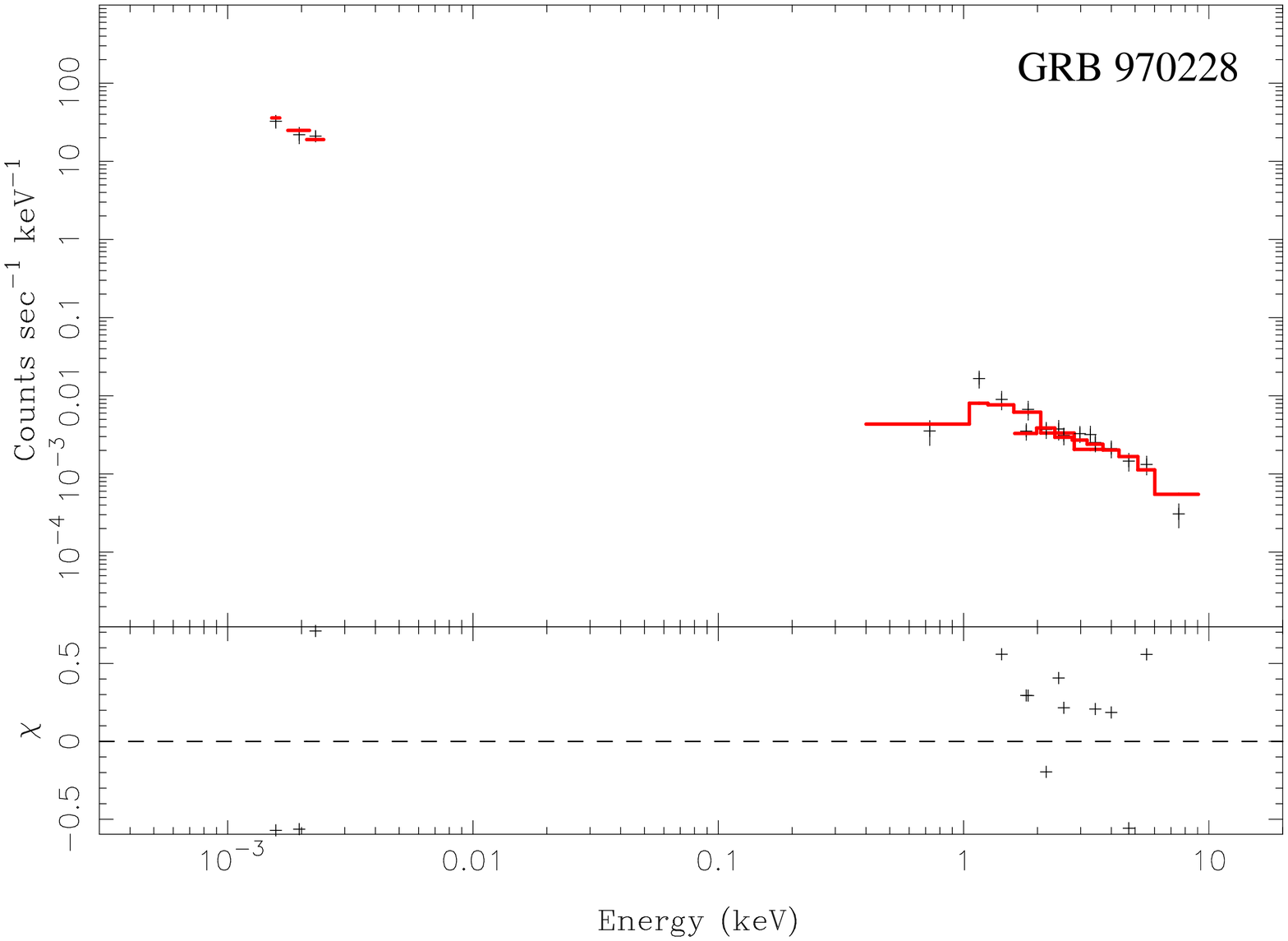}
\includegraphics[width=6cm, angle=0]{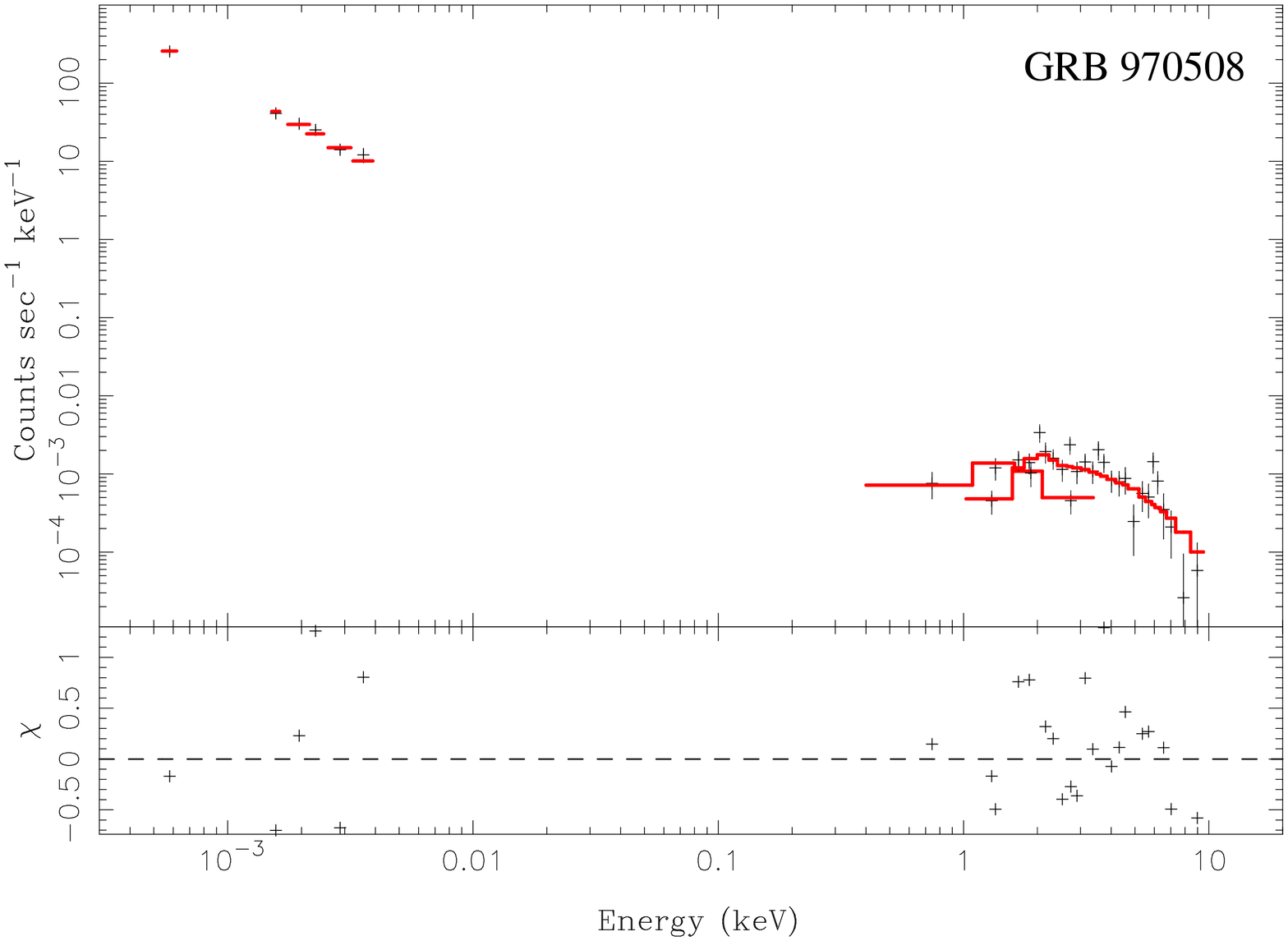}
\includegraphics[width=6cm, angle=0]{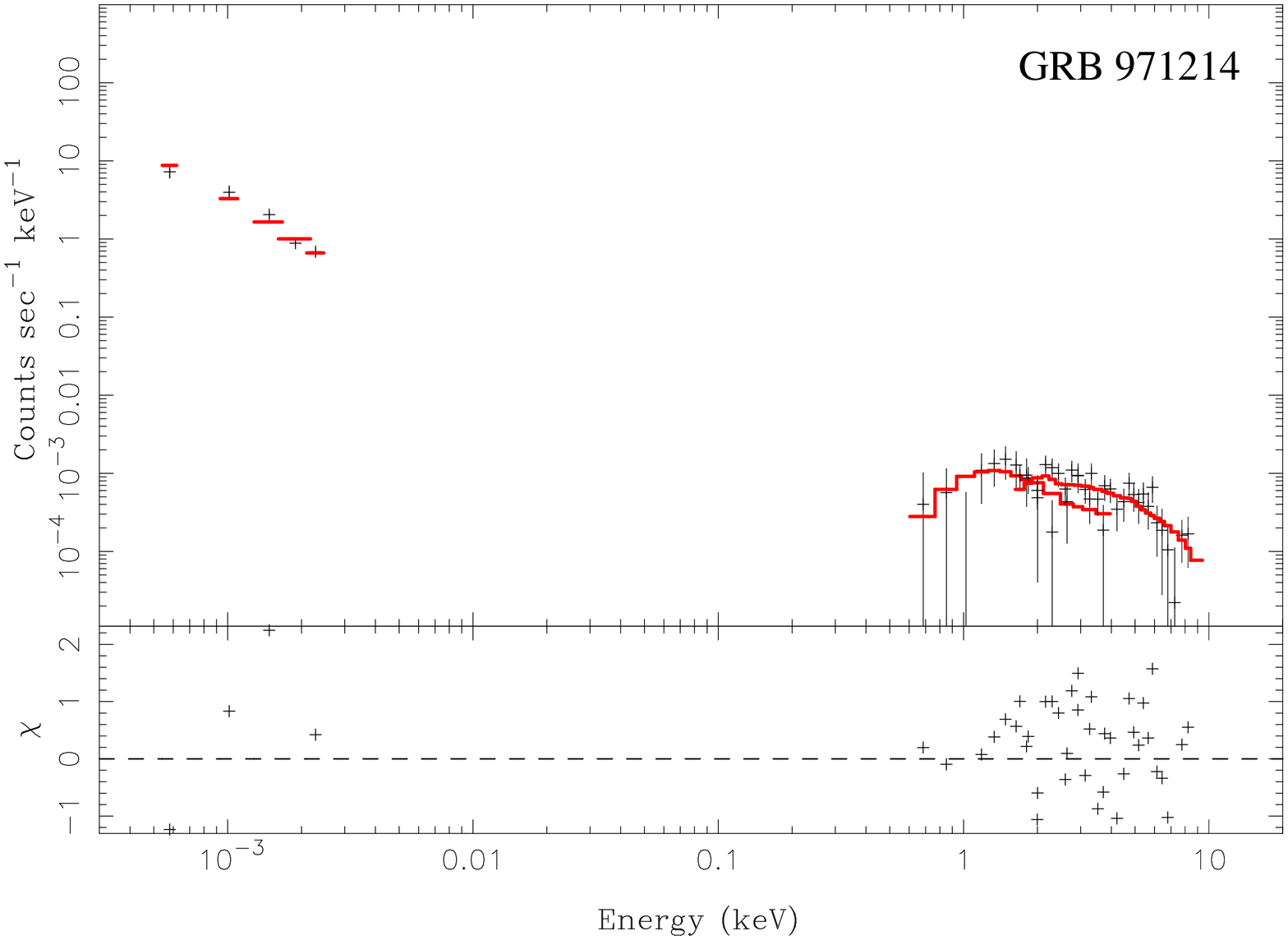}
\includegraphics[width=6cm, angle=0]{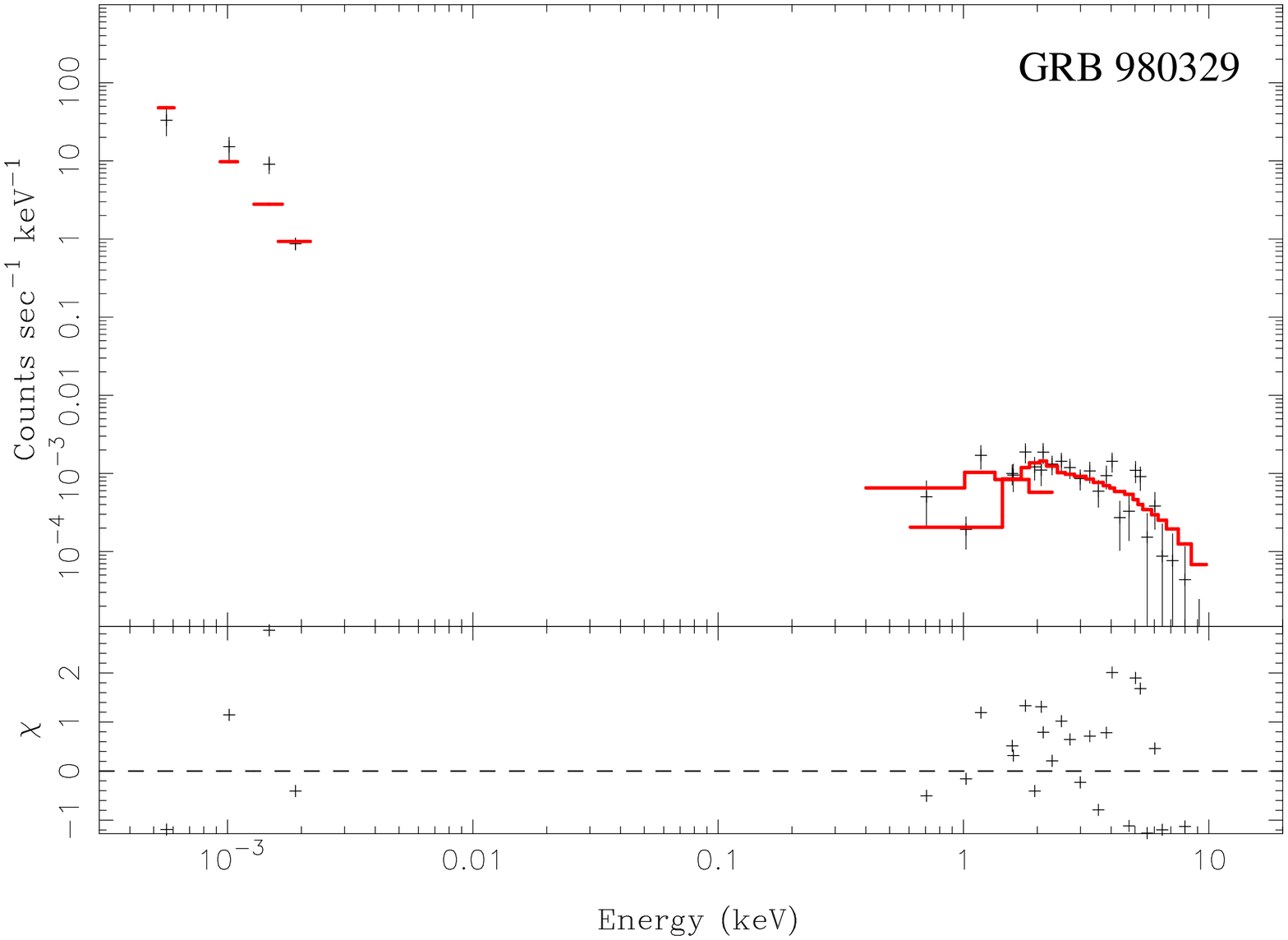}
\includegraphics[width=6cm, angle=0]{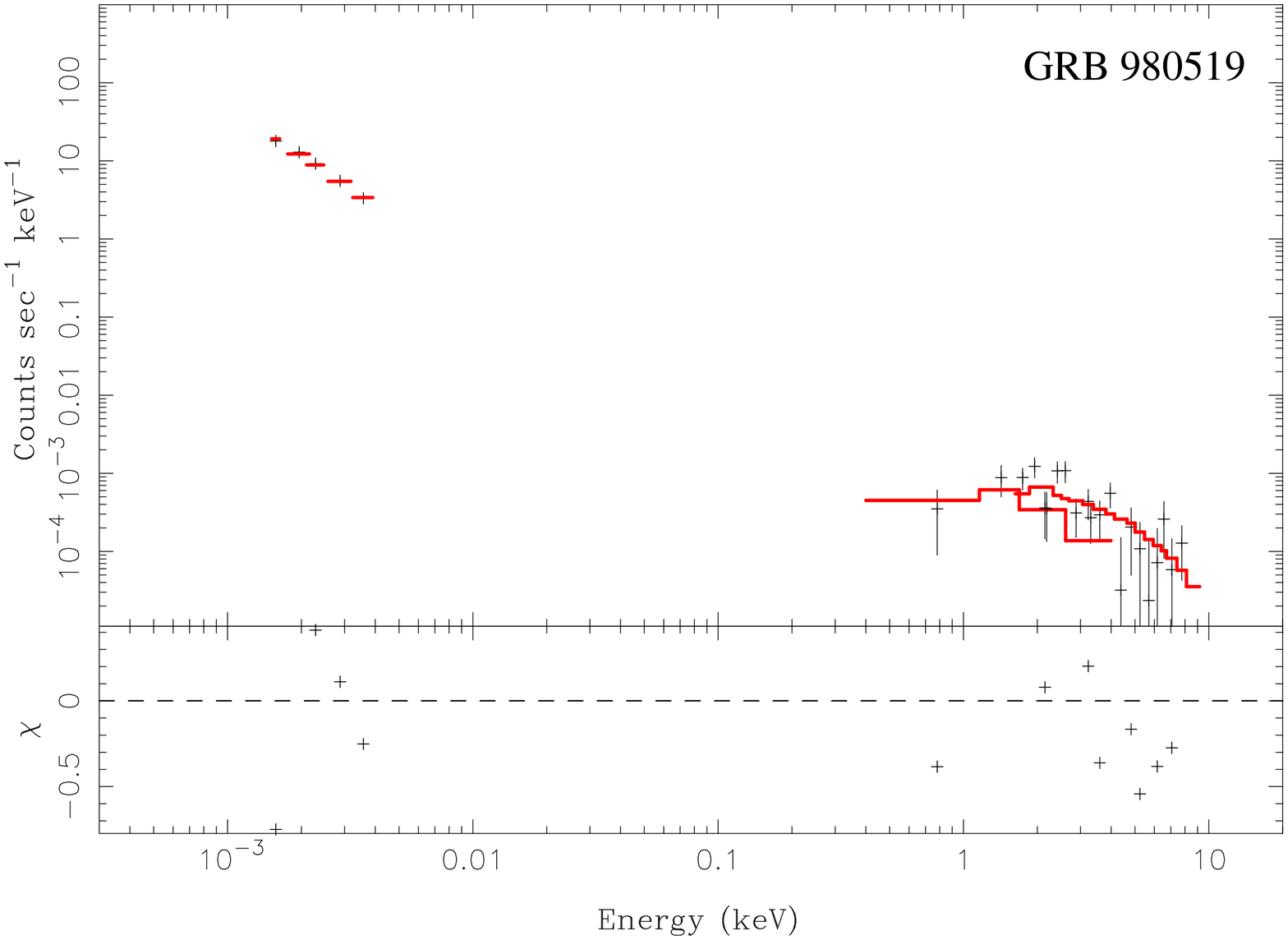}
\includegraphics[width=6cm, angle=0]{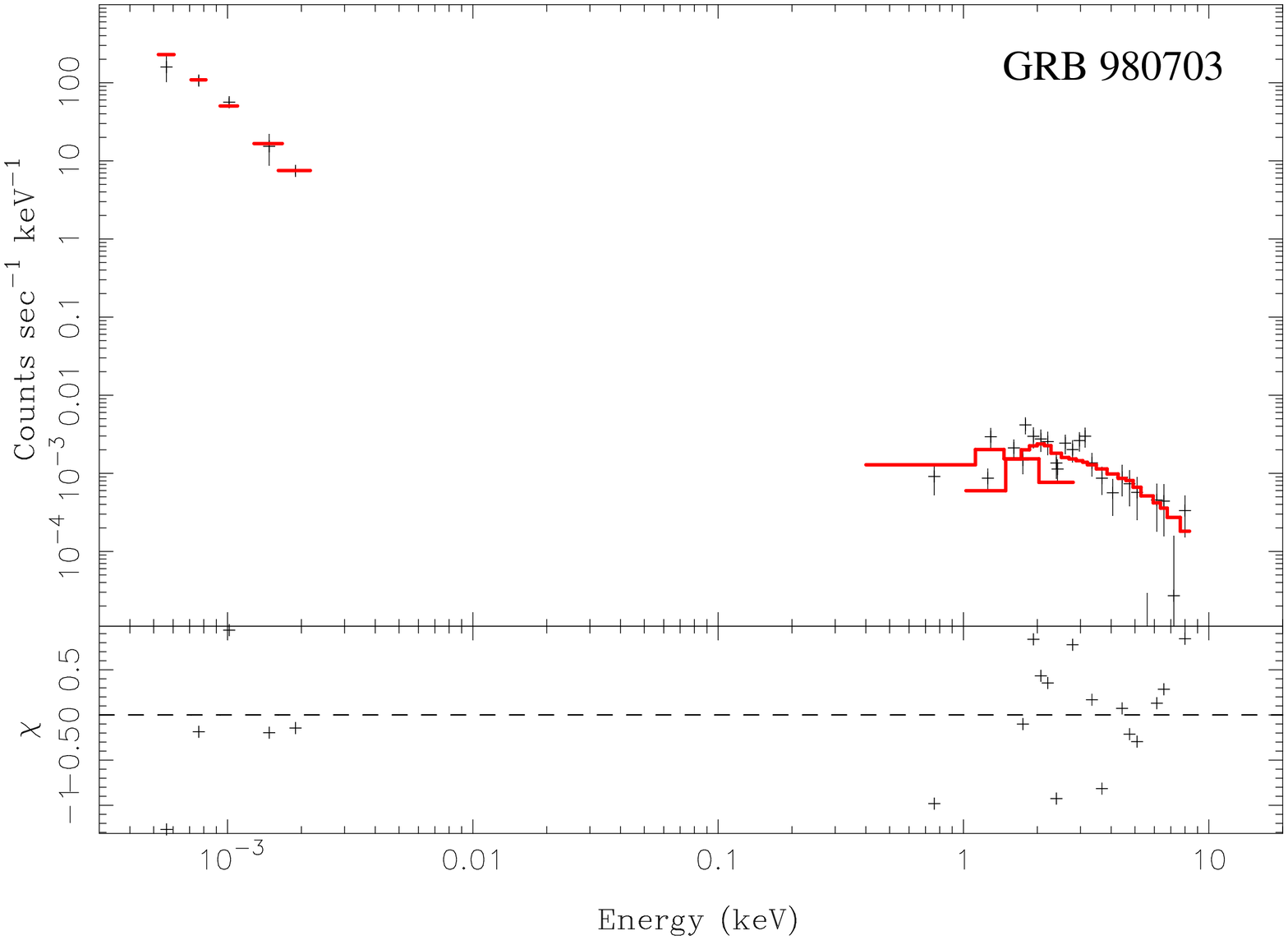}
\caption{Data (black crosses) and best fitting models (red lines, see Table~\ref{tbl-4}) for each of
  the GRBs in the sample.
 Data (nIR and optical photometry and {\it BeppoSAX} LECS and
  MECS X-ray spectra) and models are shown in count space. The bin size
  (effective bandwidth) of the optical data points can be seen in the model fits. The lower panels show the deviation from the model for individual data
points, in units of their measurement error.\label{fig2}}
\end{center}
\end{figure*}

\addtocounter{figure}{-1}
\begin{figure*}
\begin{center}
\includegraphics[width=7cm, angle=0]{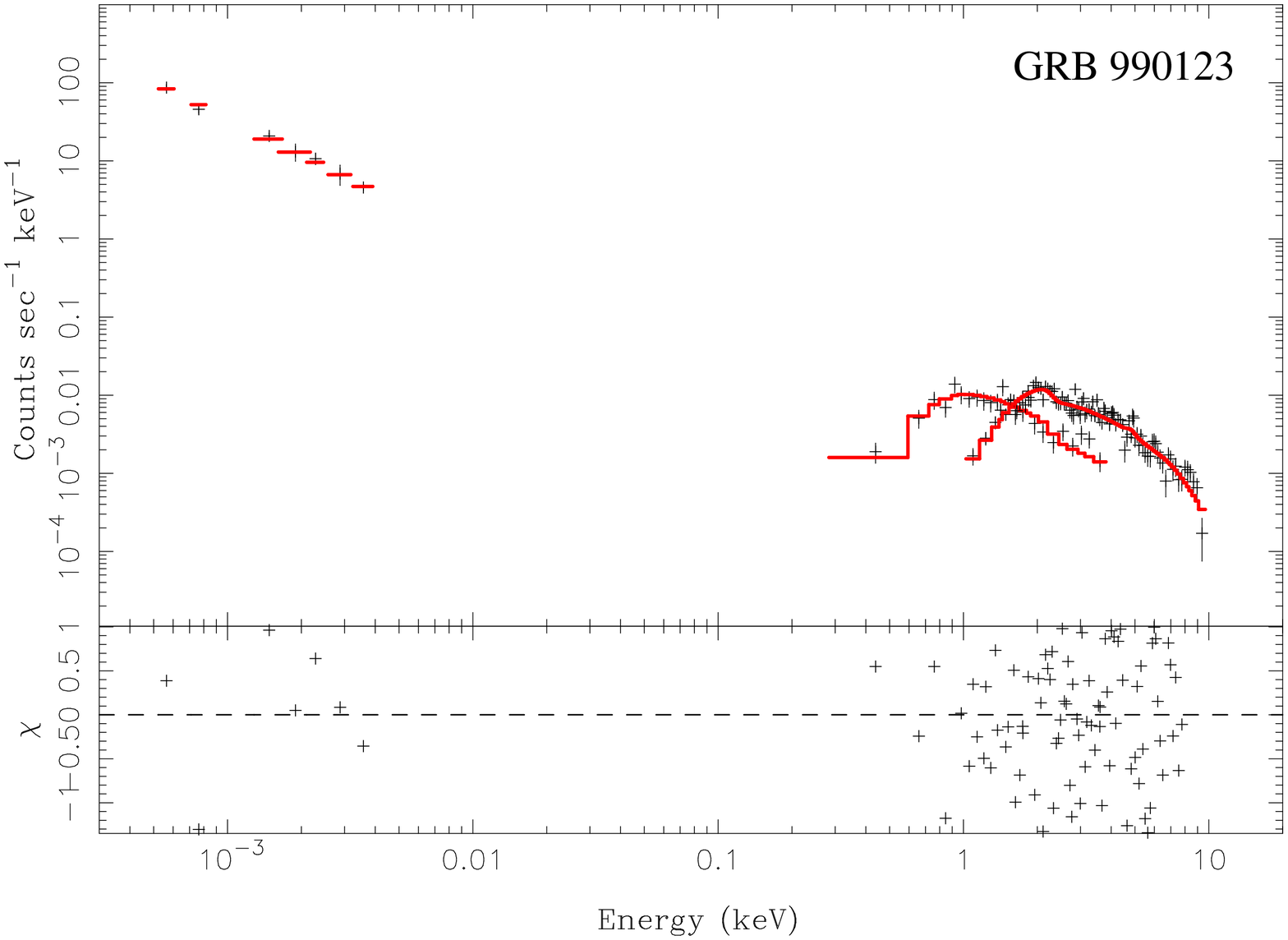}
\includegraphics[width=7cm, angle=0]{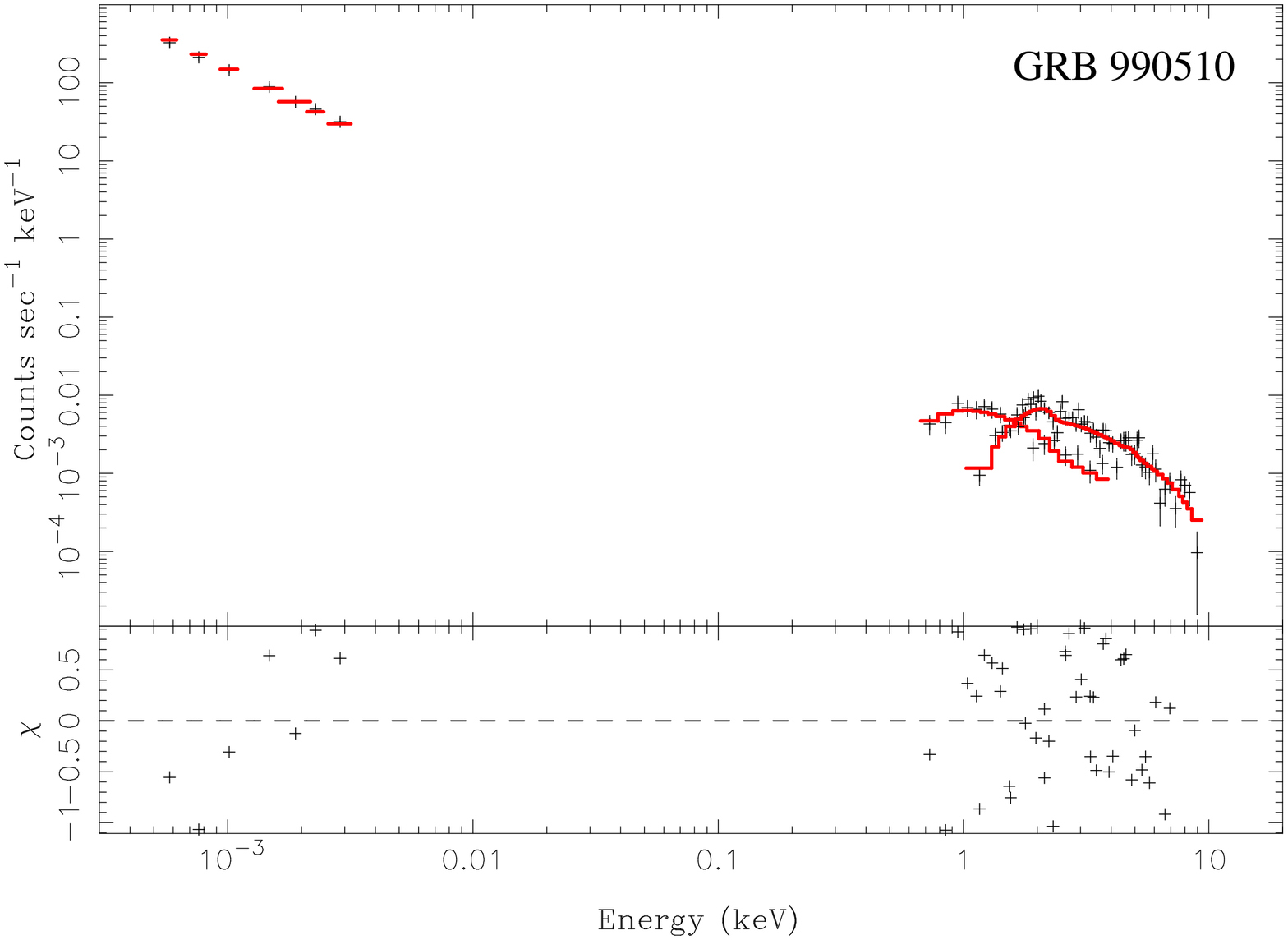}
\includegraphics[width=7cm, angle=0]{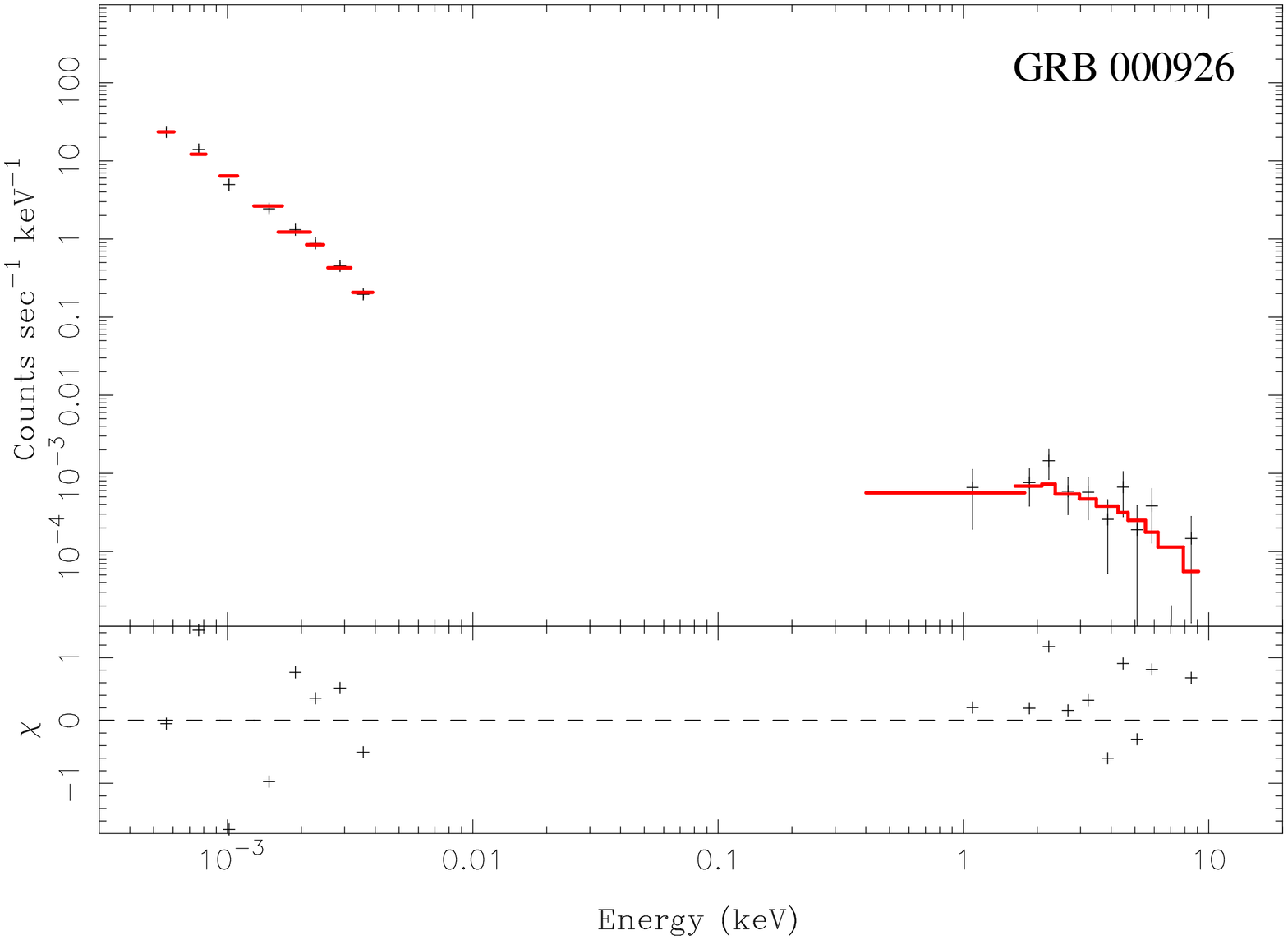}
\includegraphics[width=7cm, angle=0]{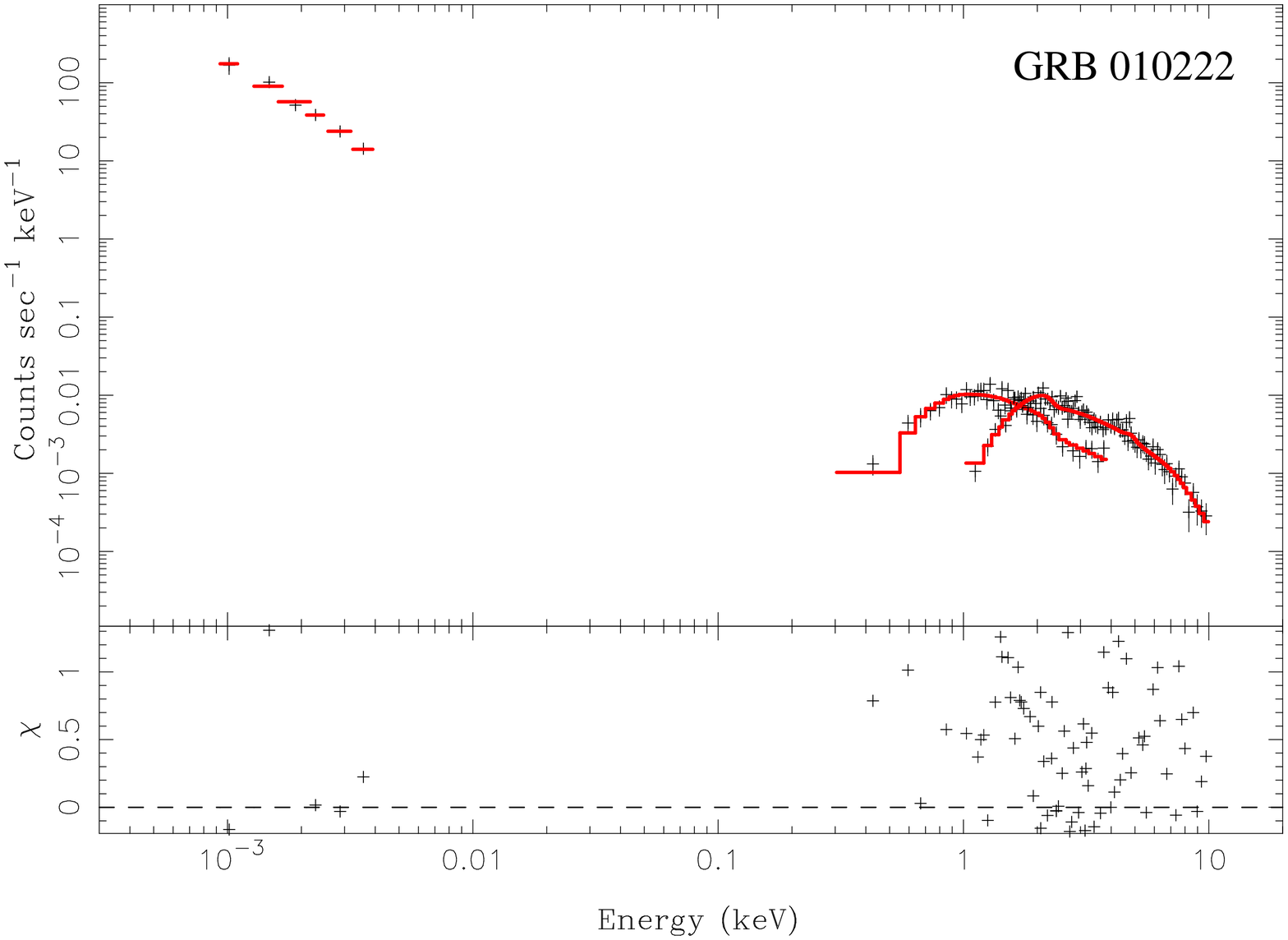}
\caption{-- {\it continued}}
\end{center}
\end{figure*}

\section{Results} 
\begin{figure*}
\begin{center}
\includegraphics[width=12cm,angle=90]{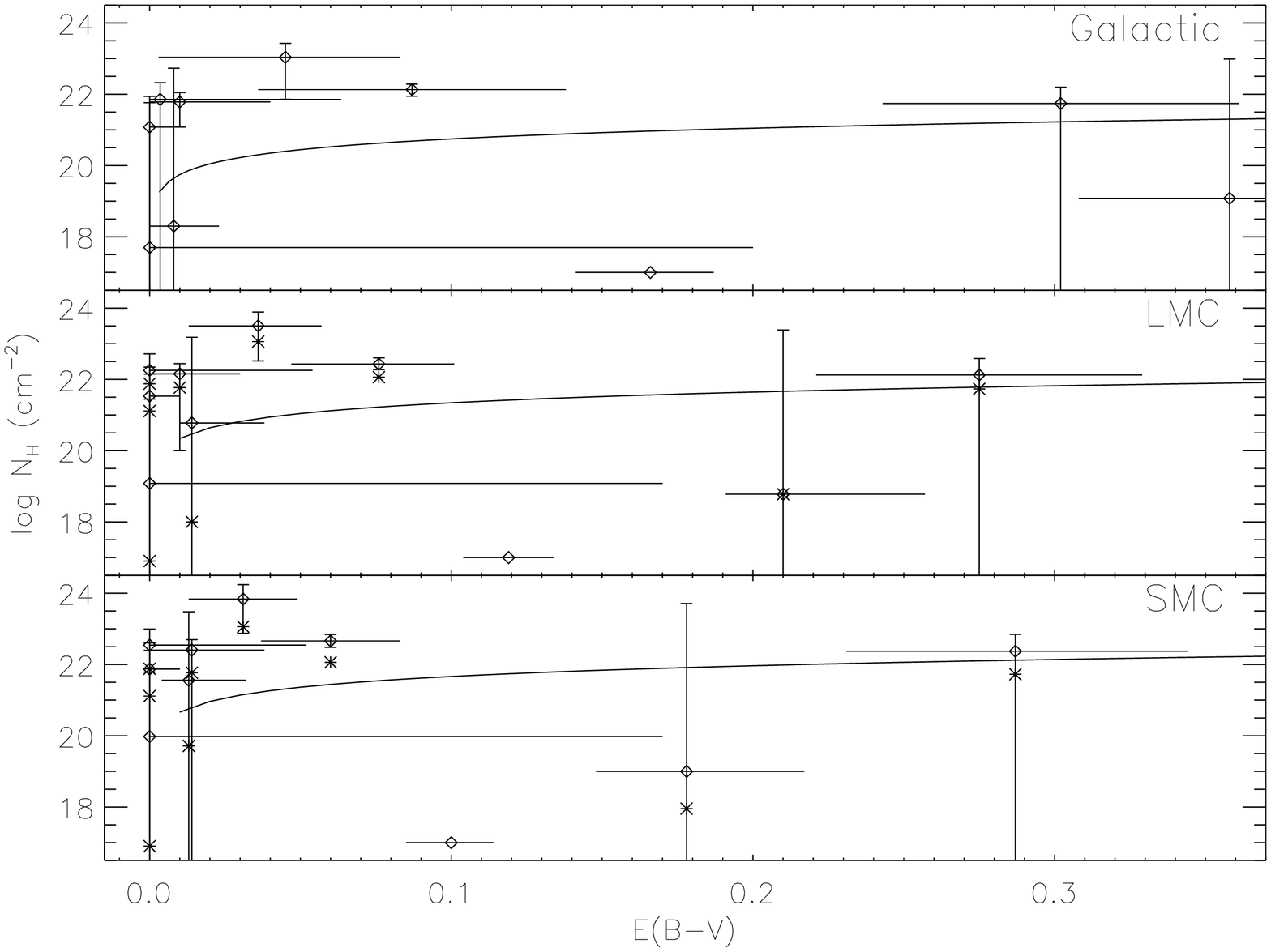}
\caption{Intrinsic absorption in optical/nIR
  ($E(B-V)$) and X-rays (log $N_{\rm H}$) measured for the GRB sample. We
  compare the measurements with three different optical
  extinction laws overlaid with solid curves: Galactic (top panel, Predehl \& Schmitt
  1995), LMC (middle panel, Koornneef 1982; see also Fitzpatrick 1985) and SMC
  (lower panel, Martin et al. 1989). Appropriate
  metallicities are adopted for LMC (1/3 Z$_{\odot}$) and SMC (1/8
  Z$_{\odot}$) calculations (diamonds), and stars mark the
  centroids of the Solar metallicity fits. For
  GRB\,000926 the data were too sparse to fit for $N_{\rm H}$, so we plot the
  $E(B-V)$ range at log $N_{\rm H}$ = 17.0 for clarity. Error bars are 90 per
  cent confidence.
\label{fig3}}
\end{center}
\end{figure*}

\begin{figure}
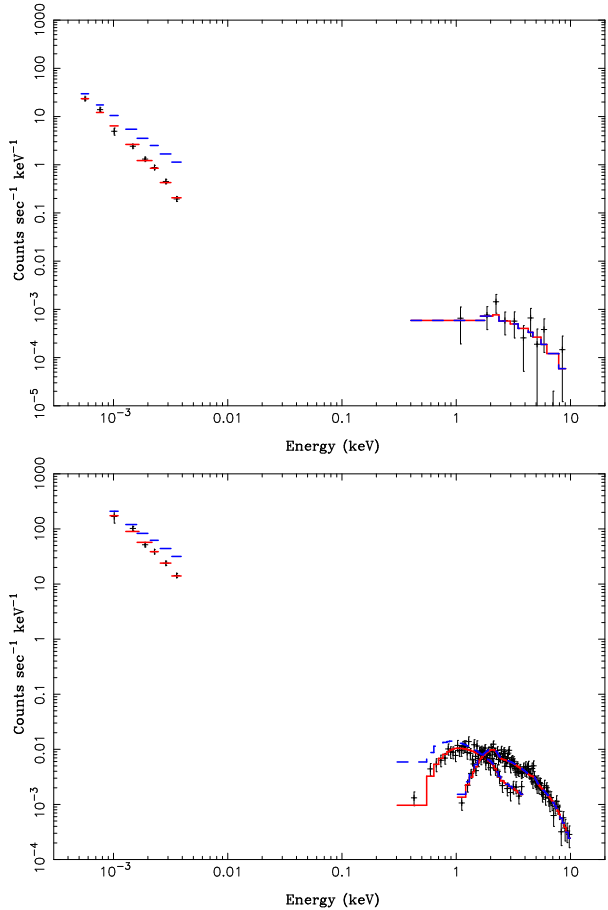

\begin{center}
\includegraphics[width=6cm, angle=-90]{f4a.ps}
\includegraphics[width=6cm, angle=-90]{f4b.ps}
\caption{Spectral energy distributions of GRBs 000926 (left) and 010222 (right) with best fitting models overlaid in red. Overlaid in blue with a dashed line is the unabsorbed source flux, demonstrating that for these two sources extinction significantly affects the observed optical to X-ray emission, in complete contrast to GRBs 970228 and 990510 where extinction in both optical and X-ray regimes are negligible.
\label{fig4}}
\end{center}
\end{figure}

The SEDs and the results of fits to the SEDs for all GRBs in the sample are
listed in Table~\ref{tbl-4}, and best fitting models are shown overlaid on the data in Figure~\ref{fig2}. Figure~\ref{fig3} shows a comparison of the absorption measurements with Galactic, LMC and SMC
gas-to-dust ratios, which we discuss in the following section. This plot has
been constructed in a number of previous works (e.g. Galama \& Wijers 2001;
Stratta et al. 2004; Kann et al. 2006; Schady et al. in preparation) but here we show
the observed distribution of $E(B-V)$ and $N_{\rm H}$ for the first time
derived simultaneously from a fit to X-ray, optical and nIR data. We find an excess in absorption above the Galactic values particularly significant in two sources: GRBs 000926 ($E(B-V)$ only) and 010222 (Figure~\ref{fig4}), whilst no significant intrinsic absorption is necessary in GRBs 970228 and 990510. The cooling break can be located in three of the afterglows: GRBs 990123, 990510 and 010222 and to all other SEDs a single power law is an adequate fit. Details are given below for each individual afterglow.

\subsection{Notes on individual sources}

\subsubsection{GRB\,970228}
No significant absorption is measured for GRB\,970228. We find only a single
power law is required and it is possible to pin down the power law slope
relatively well ($\chi^2$/dof $=$ 11.6/14).

\subsubsection{GRB\,970508}
A well-defined temporal decay slope starting 1.9 days after trigger (Zeh et al. 2006). Preceding this time there is an
increase in flux followed by an apparent flattening. The time of the X-ray
observation log midpoint occurs a little before the 1.9-day break at 1.679
days, so we have extrapolated the optical data post 1.9 days back to 1.9 days
and then assume that the evolution of the lightcurve is flat back to 1.679
days after trigger. To allow for a different behaviour before 1.9 days we
include a constant value offset in the model between the optical and X-ray
data, which we both fix at 1.0 and leave as a free parameter. The improvement
in the fits when the offset is a free parameter is somewhat marginal. An F-test indicates that the free-parameter model
is better at a 98\% level, with the offset increasing to
3. Given the uncertain extrapolation, however, we use the fits with
the offset as a free parameter in our further analysis.
The best fitting model for GRB\,970508 is a single power law with relatively
low intrinsic X-ray absorption at the level of $N_{\rm H} \sim$ 10$^{21}$
cm$^{-2}$. Kann et al. (2006) found a best fit with MW-like dust, but as we measure no significant optical extinction we cannot distinguish between different extinction laws. 

\subsubsection{GRB\,971214}
GRB\,971214 is the highest redshift source in the sample at $z$ = 3.418, and
we note that the source was faint, particularly as seen by LECS, crucial for
the low X-ray energies. We measure an optical-to-X-ray spectral index
$\alpha_{ox}$ of 0.6 and
find that a single power law is an
acceptable fit to these data. The intrinsic X-ray absorption appears to be
extremely large whilst the optical extinction is moderately large, but we note
that the high redshift of the GRB makes measurement of the X-ray extinction
more difficult. The curvature in the optical part of the spectrum has been
previously interpreted as a cooling break (Wijers \& Galama 1999) and as
extinction by either SMC-like extinction (Stratta et al. 2004) or by a presently unknown, more complex extinction law (Halpern et al. 1998; Ramaprakash et al. 1998). Whilst SMC extinction is the best fitting law of the 3 used here, it is not sufficient to accurately reproduce the shape of the optical SED, reflected in the large errors on $N_{\rm H}$ and $E(B-V)$.

\subsubsection{GRB\,980329}
In the absence of an accurate redshift determination for GRB\,980329 we adopt
the photometric redshift of $z$ = 3.6 (Jaunsen et al. 2003), hence all results must be taken with caution.
The most striking feature of this SED is the apparent flux deficit in the $R$ band,
which is present even after correction for Galactic absorption and the
probable high redshift.
We tested for the possibility that the R-band flux deficit is due to the
2175\AA\ feature in the Milky Way extinction curve by fitting the pl+MW model
whilst leaving the redshift as a free parameter lying between $z$ = 1.2 and
$z$ = 4.2 (Jaunsen et al. 2003) (we note that at $z$ = 3.6 the 2175\AA\ bump
would lie between the $I$ and $J$ bands if extinction is MW-like, which is not
observed). We obtain a fit with $\chi^2$/dof = 32.7/26 and $z$ =
1.95$^{+0.29}_{-0.31}$. The other best-fitting parameters, as listed in Table
5 for standard fits, do not differ greatly from those found in the $z$ = 3.6
fit and the optical curvature is no better fitted. We also tested for the
possibility that the R-band deficit is caused by a break in the power law, by
allowing the break energy to reside within the optical regime (and adopting
$z$ = 3.6 with SMC extinction). In this fit the break energy could not be well
constrained, and the power law slopes are not well described by the difference
of 0.5 as expected for a cooling break (the second slope is steeper), hence we
rule out this possibility. However, it is also possible that we are seeing an
$I$ band excess rather than an $R$ band deficit. Two late time $I$ band
points taken at 1--10 days since burst (Yost et al. 2002) appear not to agree
with all of the $I$ band data used here (Reichart et al. 1999), which could be
the result of overestimation of the early $I$ magnitudes, underestimation of
the later $I$ magnitudes or the occurrence of color evolution.
Among the models applied to the whole GRB sample here, a single power law unabsorbed in the X-rays is the best fit, with a moderate $E(B-V)$ of 0.178 and a slight preference for SMC extinction.

\subsubsection{GRB\,980519}
We caution that all the results are based on a redshift estimate equal to the
mean of the sample spectroscopic redshifts of $z$ = 1.58. A few attempts to
constrain the redshift have not been very precise, amounting to 0.5$<z<$3.6 or
1.5$<z<$3.6 (lower limits from Jaunsen et al. 2003 and upper limit from the
fact that we detect a $U$ band counterpart). We find that a single power law with a small $E(B-V)$ and X-ray absorption consistent with zero but with a large error is sufficient to model this afterglow. We note that the Galactic X-ray extinction towards GRB\,980519 is the highest for this sample which, together with the lack of known redshift hampers a good measurement of $N_{\rm H}$ for this source.

\subsubsection{GRB\,980703} 
A single power law provides an acceptable fit to the spectrum, when absorbed by a large amount at both X-ray and optical/UV wavelengths. In this afterglow we measure the largest $E(B-V)$ value in the sample of 0.29$\pm$0.05 or $A_{V}$$\sim$0.85 at 1.33 days after burst assuming SMC extinction (which is marginally preferred). This is consistent with the value found by Bloom et al. (1998b) of $0.9\pm 0.2$, 5.3 days after the burst. 

A great deal of work has been done on the host galaxy properties of this burst
because the host is bright, with optical extinction measurements by 5
groups. There is a hint that the optical extinction may be decreasing with
time (see e.g. Holland et al. 2001) since measurements of $A_{V}$ at different
times are inconsistent: $A_{V}$$\sim$2.2 at 0.9 days, Castro-Tirado
et al. 1999; $A_{V}$$\sim$1.5 at 1.2 days, Vreeswijk et al. 1999;
$A_{V}$$\sim$0.3 at 4.4 days, Djorgovski et al. 1998; $\sim$0.9 at at 5.3
days, Bloom et al. 1998b, and there appears to be a discrepancy between the
measured optical spectral and temporal slopes when assuming $A_{V}$ is
constant. However, we note that the optical spectral slope was taken to be
$\beta_{\rm OA}$ = -2.71$\pm$0.12 from Vreeswijk et al. (1999), and in this
study we obtain a lower value of $\beta_{\rm OA}$ = -0.92$\pm$0.03 which would
be completely consistent with $\beta_{\rm OX}$ = (1+2$\alpha$)/3 = 0.9 using
$\alpha$ = 0.85 from Zeh et al. (2006, noting their $\alpha$ value has a very large associated error - Table~\ref{tbl-1}). We use the Vreeswijk et al. (1999) optical data here and scale it from 1.2 days to 1.33 days after trigger. Combining the optical and X-ray data when fitting provides us with a different estimate for the extinction than was obtained by Vreeswijk et al. for the optical data alone. We note that we have made only a minimal extrapolation of the original optical data used in this analysis, from $\sim$1.2 days to 1.3 days after trigger. 

\subsubsection{GRB\,990123}
GRB\,990123 does not have significant X-ray absorption above the Galactic value, and the optical extinction is consistent with zero. We set an upper limit to the latter which is ten times lower than the value found by Savaglio et al. (2003) from fits to Zn~II and H~I in the optical spectrum. 
A single power law fit to both optical and X-rays results in a spectral slope
of $\beta$=0.61$\pm0.01$ at 1.24 days since burst, comparable to the
$\beta_{\rm OX}$=0.67$\pm$0.02 at about the same time since the burst found by
Galama et al. (1999). The latter authors note that the cooling break must lie
at or above X-ray frequencies at that time. However, we find an improved fit
with a broken power law, constraining the cooling break to 0.15 $<\nu_c<$ 2.4
keV, within the X-ray spectrum, also found by Stratta et al. (2004), Corsi et
al. (2005) and Maiorano et al. (2005). We note, however, that the X-ray
spectrum comprises flux accumulated between 0.2 and 2.6 days since burst,
including the proposed jet break time of 2 days (Table~\ref{tbl-2}). We tested for the possibility that the offset between optical and X-ray data was incorrect, but this made very little difference to our overall goodness of fit. It is possible that the cooling break has entered the X-ray band during these observations, since this break is expected to decrease in frequency with time. If this is the cooling break, the spectral slopes above and below the break differ by the expected factor of 0.5 when left free.
We therefore took only the data from the first X-ray observation, before the jet break (Zeh et al. 2006), together with scaled optical/nIR data, which has a log midpoint of $\sim$0.7 days since trigger. We repeated all our fits. We find acceptable fits again only with the broken power law model, with the break energy lying somewhere between the optical and $\sim$2 keV (preferring central values just below X-ray frequencies, in line with our X-ray only analysis in which a single power law is a good fit to all spectra). 
We note that an excess of flux at high energies (as seen by the {\it BeppoSAX}
PDS instrument not used here) is reported by Corsi et al. (2005). They
attribute this to an Inverse Compton component (though we note that this
remains inconsistent with the radio data, Kulkarni et al. 1999b). Since the X-ray spectrum is adequately fit with a single power law we assume that the tail of any such component is not significant below 10 keV.

\subsubsection{GRB\,990510}
Fits with Galactic, LMC- and SMC-like extinctions show that $E(B-V)$ is very
low in this source and we can only determine upper limits. The low amount of
extinction makes it impossible to pin down the extinction curve shape, hence a
fit using Galactic-like extinction is sufficient here and LMC- and SMC-like
extinctions give similar results. There is considerable improvement in the
$\chi^2$ when allowing for a break in the power law, noted by previous
authors, which we find is located at 0.016--0.020 keV at $\sim$1.06 days since
burst (of the order of the value of $\sim$0.029 keV at $\sim$1 day since burst given by Pian et al. 2001). The slope change is as expected for a cooling break in the slow cooling regime when leaving both power law slopes free.
Our measured spectral slope in the optical regime agrees well with that
measured in the Very Large Telescope (VLT) spectra of $\beta$ = 0.6$\pm$0.1 by Vreeswijk et al. (2001).

We note that Kuulkers et al. (2000) analysed these X-ray data in several time bins and found no spectral evolution, hence the cooling break remains outside the X-ray frequencies during the observations.
We constrain the optical extinction to be $E(B-V)$ $\le$ 0.01, and the X-ray
equivalent hydrogen column to be $\le$ 0.15$\times$10$^{22}$ cm$^{-2}$ for SMC
extinction and $\le$ 0.87$\times$10$^{22}$ cm$^{-2}$ for MW extinction. From
the X-ray spectrum alone a higher column of $N_{\rm H}$ = 2.1 $\pm$
0.6$\times$10$^{21}$ cm$^{-2}$ was measured by Kuulkers et al. (2000). Optical
spectra have provided a lower limit on the amount of neutral hydrogen towards
GRB\,990510 of log $N$(H~I) $\ge$ 19.7 cm$^{-2}$ (Vreeswijk et
al. 2001). These authors obtain an approximate estimate for the metallicity
from the optical spectra using Fe/H, and find 12+ log [Fe/H] = -1.5 $\pm$ 0.5
or 0.01--0.1 times the Solar value. This range approximately covers the
metallicity of both the LMC (0.33, Pei 1992) and the SMC (0.125, Pei 1992) so
we also fitted the data with these two metallicities. As there is no
substantial absorption observed, the fits do not change significantly.

\subsubsection{GRB\,000926}
The optical and IR SED of this burst is very well sampled, but the X-ray
afterglow was very faint at the time of observation. For this reason these
X-ray data have been grouped to have a maximum of only 10 counts per bin, and
strictly speaking this means Gaussian statistics should be treated with caution. However, we do use the $\chi^2$ statistic as a goodness of fit for comparison with the rest of the sample.
Given the poor quality of the X-ray data, and the lack of sufficient counts in
low X-ray energy bins, we have fixed the X-ray column density at zero and fit
only for the optical extinction. We find a large amount of intrinsic
extinction is necessary to describe the flux deficit with respect to a single
power law, consistent with the value found by most other studies (Stratta et
al. 2004; Fynbo et al. 2001; Harrison et al. 2001; Price et al. 2001). A large
extinction was also derived from an optical spectrum by Savaglio et
al. (2003), and the $A_{V}$ found in that study is approximately twice the
value found here. However, Savaglio et al. use the spectral line measurements
to first fit for the depletion pattern and then infer an extinction. Harrison
et al. (2001) interpreted the optical flux deficit as indicating that a
significant fraction of the X-ray flux was in fact Inverse Compton emission -
later also suggested for GRB\,990123 (see above).
This is the only afterglow for which an LMC extinction law is (marginally)
preferred. Fynbo et al. (2001) report a tentative H~I column density
measurement of $N$(H~I) $\sim$2$\times$10$^{21}$ cm$^{-2}$ which leads to a
relatively high metallicity with [Zn/H] = -0.13. This metallicity is between
the LMC and the MW values, which may explain the preference here for LMC-like
extinction if no 2175\AA\ bump is present.

\subsubsection{GRB\,010222}
A good dataset for GRB\,010222 allows the spectral properties to be well
constrained. We find that whilst a single power law is a reasonable fit to
these data, a broken power law significantly improves the fit. The break
energy lies around 0.01 keV, above the frequency of the last optical band in
our SED. Optical extinction is clearly non-zero, with $E(B-V)$ = 0.06$\pm$0.02,
consistent with that found by Lee et al. (2001) for an SMC extinction law but
about three times lower than that inferred from the spectral lines by Savaglio
et al. (2003). X-ray absorption is also required with an effective hydrogen
column of $N_{\rm H}$ = 1.15$^{+0.54}_{-0.39}$ $\times$ 10$^{22}$ cm$^{-2}$. 

Panaitescu \& Kumar (2002) in their fits to the multiwavelength afterglow of
this source
find significant reddening of the optical spectrum of $A_V = 0.21$ with an SMC
extinction curve, which is consistent with our value. But they find a large
fitting error and attribute this to 8 outlying points suggesting that either
some reported observations have underestimated uncertainties or there are
short timescale fluctuations in the afterglow emission (Cowsik et
al. 2001). Their jet model requires the cooling break to pass through the
X-ray band at about 1 day, which they find incompatible with their
observations. Our analysis places the cooling break at
optical/UV wavelengths at 1.51 days since burst.

\subsection{Comparison with previous studies}
We can compare our results directly with those of previous studies of
samples overlapping with this {\it BeppoSAX} subsample. In general, we are finding
similar central values for extinction as all previous studies, and are
improving upon the uncertainties, fitting all afterglows in the same consistent manner allowing for direct comparison.

Galama \& Wijers (2001) performed the first systematic study of line-of-sight $N_{\rm H}$ and $E(B-V)$ with
a sample of 8 afterglows, consisting of all but
the two most recent bursts in our sample. From fits to the X-ray spectra they found intrinsic $N_{\rm H}$ amounting to
10$^{22}$-10$^{23}$ cm$^{-2}$, ruling out the possibility that some hosts have no X-ray
column at all. They noted that these values lie in the range of Galactic
giant molecular clouds (estimating cloud sizes of 10--30 pc) - a conclusion also recently arrived at when including Swift bursts
(Campana et al. 2006a) and when measuring H~I from damped Lyman alpha absorption in GRB optical afterglow
spectra (Jakobsson et al. 2006a).
They used a simple extinction law $A_V \propto \nu$ with a smoothly broken power
law. Comparing their dust to gas ratios with that of the Milky Way they obtain
an optical extinction 10--100 times smaller than expected. Their finding of generally low $A_{V}$ was
attributed to previously predicted dust destruction by the GRB. Of the bursts
common to both their and our samples and having a known redshift we obtain
consistent extinction values within the 2$\sigma$ uncertainties in all cases except for GRB\,971214 for which the
two optical extinctions are only consistent at the 3$\sigma$ level, our central value
lying four times lower than the Galama \& Wijers measurement. We obtain
smaller uncertainties in our extinction measurements in all cases except for
the X-ray column of GRB\,980703.

Stratta et al. (2004) have also measured $N_{\rm H}$ and A$_{V_r}$ in the optical and
X-ray data seperately, and later plotted the combined data in flux space (after
assuming the X-ray model to be correct) in order to judge the position of the
cooling break. To derive optical spectral slopes, Stratta et
al. first adopt the $p$-value derived from the X-ray fits (the input electron energy index, see Paper II
for our fits for this parameter), use this to fix the
optical spectral index and fit for $A_{V}$. Our results for $E(B-V)$ with
SMC-like extinction are consistent in all cases to within the 90 \%
confidence limits, adopting the cooling break positions found from this work
if known (the values for GRB\,970508 are only consistent if $\nu_o>\nu_c$).
Even for the two GRBs for which we have assumed different redshifts we find
consistency in extinction estimates (Stratta
et al. adopt $z$ = 1 for both 980329 and 980519). In 7 of the 10 cases we
derive a better constrained value or upper limit to the extinction. The X-ray
absorbing columns we measure are also generally consistent with those found
by Stratta et al. in a fit to the X-ray data only (these are of course the
same data used in our study, except for differences in number of observations
combined for some sources, and we include GRB\,000926 as well). Our
method obtains improved estimates for $N_{\rm H}$ for 7 of the 9 sources
common to both studies. For GRB\,990510 Stratta et al. find a value higher
than our derived $N_{\rm H}$ using
Solar metallicity, and for GRB\,980329 our result for $N_{\rm H}$ is less accurate at the 90
\% confidence level, though we note the different redshifts assumed hence
direct comparison is not possible.

Kann et al. (2006) fitted only the optical SEDs for a sample of pre-Swift
bursts including 8 analysed here (not including GRBs 970228 and 980329 and
using $z$ = 1.5 for GRB\,980519). We find similar central values of $A_V$ and
improve upon their mean error by a reduction of 5--10 \%.  
Values disagree at the 90 \% confidence level only for GRB\,970508,
where we find an upper limit to the extinction which is 2.4 times lower than
the lower limit of Kann et al. but would be consistent with their estimate at
the 3$\sigma$ level.

\section{Discussion} 

\begin{figure}
\begin{center}
\includegraphics[width=7cm, angle=0]{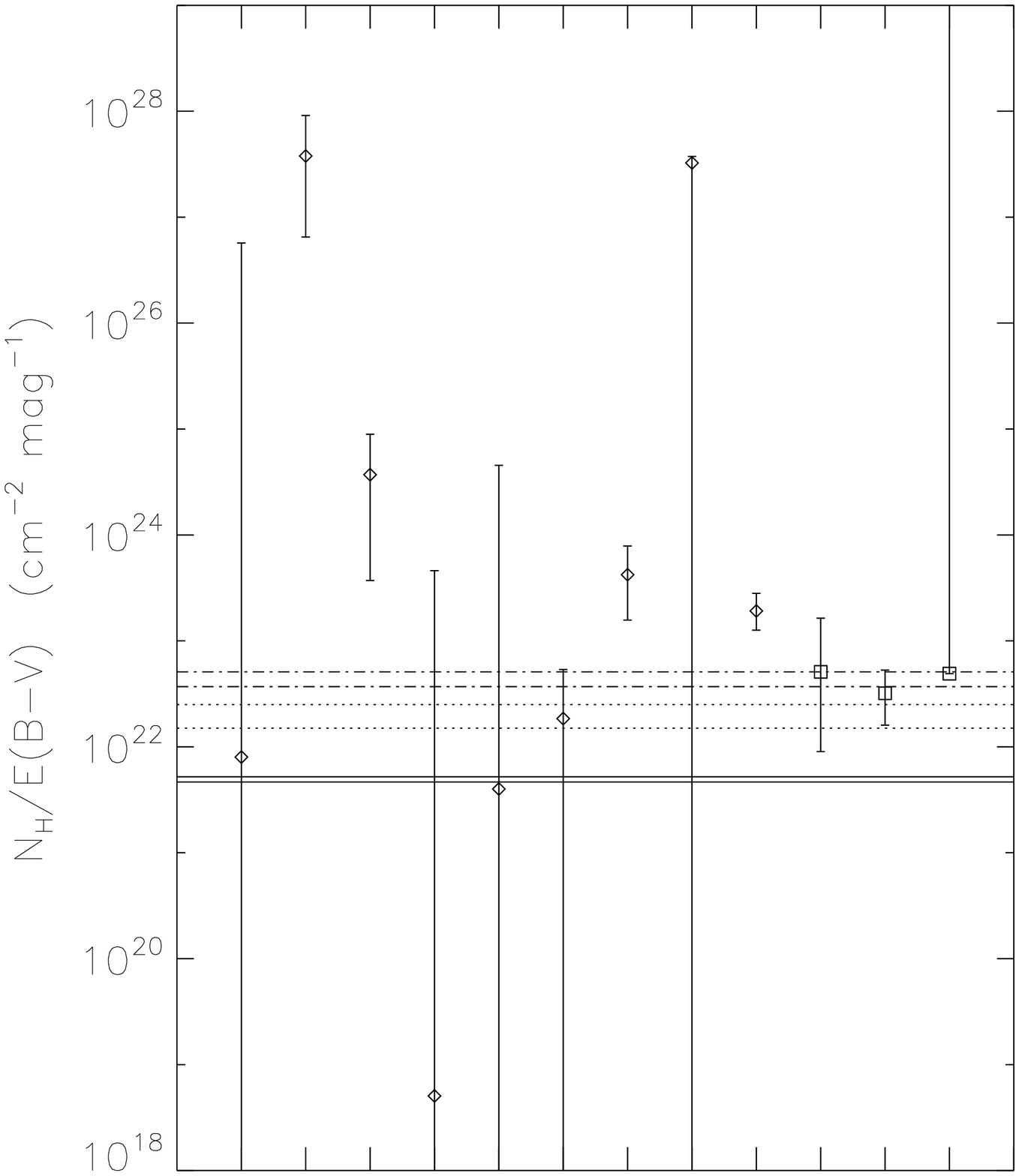}
\caption{Gas-to-dust ratios, $N_{\rm H}$/$E(B-V)$, derived from best fits to
  the SEDs assuming SMC metallicity, plotted for each sample GRB in date order from
  left to right (GRB\,970228 first), excluding GRB\,000926 where we did not
  fit for
  X-ray column density. We include the ratios for 3 GRBs taken from the literature for
  comparison (square data points): GRBs 000301c (Jensen et al. 2001), 000926
  (Fynbo et al. 2001) and the lower limit on 020124 (Hjorth et al. 2003). The
  solid, dotted and dot-dashed lines show the measured gas-to-dust ratios and
  their errors for the Milky Way (Diplas \& Savage 1994), LMC and SMC
  (Koornneef 1982; Bouchet et al. 1985). Error bars are at the 90 \% confidence
  level.\label{fig5}}
\end{center}
\end{figure}

For half the afterglows the best-fitting model to the SED includes SMC-like
extinction (as opposed to LMC or MW) and in one case LMC-like extinction. In
no cases is there a preference for MW-like extinction. We are sensitive to the
2175\AA\ bump (MW) in the redshift range $z$ = 0.46--9.9, covering all our
selected GRBs, but clearly we do not detect any such feature. 
We find a wide spread in
central values for the gas-to-dust ratios, and for 4 GRBs the gas-to-dust
ratios are formally
inconsistent with MW, LMC and SMC values at the 90 \% confidence limit
assuming the SMC metallicity (Figure~\ref{fig5}). In these 4 cases the ratio is several orders of magnitude higher than
the SMC value of 4.4$\pm$0.7$\times$10$^{22}$ cm$^{-2}$ mag$^{-1}$ (Koornneef
1982; Bouchet et al. 1985) and must mean that either gas-to-dust ratios in
galaxies can span a far larger range than thought from the study of local
galaxies, or the ratios are disproportionate in GRB hosts because the dust is
destroyed by some mechanisms (likely the GRB jet), or that the lines of sight
we probe through GRBs tend to be very gas-rich or dust-poor compared with
random lines of sight through galaxies. Finally, a dust grain size
distribution which is markedly different than considered here may also affect
these ratios.

In fact, a recent study has shown that for the LMC the former is true.
A recent observation of four core-collapse supernova remnants (SNRs) in the LMC with Spitzer has
shown IR emission associated with the supernova blastwave (Williams et
al. 2006). This is interpreted as dust with an LMC-like grain size
distribution which has been collisionally heated by the X-ray emitting
plasma. The observations require that some fraction of the small dust grains
has been destroyed by sputtering by high energy ions in fast shocks. Dust
destruction is known to occur in SNR shock fronts (Jones 2004), and we will
return to this issue later in the section. The derived
gas-to-dust ratios are several
times higher than the LMC ratio, as we see in the line-of-sight
measurements of GRBs and has been observed in other types of supernova (e.g. Borkowski
et al. 2006), the cause of which is not known.

We measure a large amount of intrinsic absorption in some of the sample
(Figure~\ref{fig4}), and can state that absorption is insignificant in others, as seen for
example in the contrast between GRBs 010222 and 990510. We have tested for the possibility that the Galactic column density in
the GRB direction is affecting our intrinsic column derivations by plotting
$N_{\rm H,int}$ vs. $N_{\rm H,Gal}$, and found no correlation. We assumed
therefore that
all the measured absorption lies within the host galaxy. We note that a few of these
afterglows have spectra: for GRB\,990510 no intervening systems were clearly
identified, an intervening system is measured at 168 km s$^{-1}$ from the
host redshift for 000926 (assumed to lie within the host galaxy, Castro et
al. 2003), 2 intervening systems are found for GRB 010222 (Jha et al. 2001;
Mirabal et al. 2002) and at the host redshift there are two components
separated by 119 km s$^{-1}$ (Mirabal et al. 2002). None of these intervening
systems are close enough to us as the observer to significantly affect
measurement of the intrinsic host extinction. 

\subsection{Approaches to measuring absorption in the host galaxies}
In this study we provided a thorough, uniform study of both optical and X-ray
extinction along the lines of sight towards a sample of 10 GRB afterglows. The
well known spectral shape and relative brightness of the afterglow emission
make GRB afterglows a powerful line-of-sight probe of high redshift
extinction.
This is one of several approaches to measuring absorption in GRB host galaxies.
One can globally divide the studies of extinction in the field of GRBs in two categories: line-of-sight 
extinction studies and studies of extinction of the integrated host galaxy or parts thereof.  

\subsubsection{Line-of-sight studies}
Line-of-sight studies generally involve fitting the afterglow spectral energy distributions
in optical and/or X-rays with template extinction models (i.e. MW, LMC, SMC or
more parametrized models) as we have done here. The standard conversions between X-ray extinction and optical extinction for the Milky Way and 
the two Magellanic Clouds are generally in disagreement with the
column densities measured through this method, but these skewed gas-to-dust ratios are
also being found in other astrophysical situations as discussed above for SNRs
in the LMC, and the destruction of dust can go some way to alleviating the mismatch. 

A further way to probe line-of-sight extinction properties is through optical
spectroscopy of the afterglow. In
this case the careful measurement of column densities of heavy elements can be used to study
the dust depletion pattern along the line of sight (e.g. Savaglio \& Fall 2004).
The measured metal column densities in combination with the best-fitting depletion pattern and
the empirically determined conversion between $A_V$ and the dust column, can provide a 
prediction of the dust extinction along the line of sight to a GRB (for a detailed explanation see
e.g. Savaglio \& Fall 2004). Savaglio \& Fall (2004) show that the extinction derived from the
dust depletion method is significantly higher than the value derived from direct fitting to
the continuum of the afterglow spectrum, e.g. for GRB\,020813 they find an overestimation by at least five times. We show that this conclusion holds when fitting
the afterglow continuum emission over a much larger wavelength range and can
quantify that overestimation factor for GRBs 990123, 000926 and 010222 as
approximately $\gtrsim 11$, $\gtrsim 2$ and $\gtrsim 3$ times overestimated respectively. 
The reasons for this apparent discrepancy may be two-fold: firstly the fitted
extinction profiles to the afterglow SED are likely poor approximations to the
true extinction profile, and secondly the host dust depletion chemistry may well
differ from the Milky Way chemistry. 
In addition, the GRB or afterglow may preferentially
destroy small dust grains, skewing the extinction profile towards larger grains, resulting in 
a "grey" extinction curve. This would alter the derived extinction from SED fitting, and possibly bring 
estimates from dust depletion methods and SED fitting closer
together. Whist a grey extinction curve was the best fitting extinction curve
to GRB\,020405 (Stratta et al. 2005), we note that grey extinction
curves have been fitted to samples of afterglow SEDs with no conclusive improvement in
fit (e.g. Stratta et al. 2004).

\subsubsection{Integrated host galaxy studies}
One can also study the extinction properties
of host galaxy as a whole, and there are again several methods to do this. 
Whilst it has been shown that GRBs occur in
starforming regions in the host (Bloom et al. 2002; Fruchter et al. 2006),
many host galaxies are small and mixing timescales may be short, enabling
global properties to be measured. The host and afterglow have similar,
moderate reddening in GRB\,000418 (Gorosabel et al. 2003a) which is taken as
evidence that the ISM is well mixed. But more 
extreme values of reddening are also seen, such as the extremely red afterglow and host of
GRB\,030115, in which the host is an Extremely Red Object (ERO, Levan et
al.~2006).

One of the most common methods of integrated host galaxy studies is fitting of the
broadband optical and near-infrared SEDs of the hosts themselves (e.g. Christensen et al. 2004). Galaxy templates can be fit to the data, using photometric redshift programmes such as HyperZ (Bolzonella et al.~2000) providing values for the photometric redshift, the age of the dominant stellar population and the extinction, by fitting a series of galaxy templates. The extinction measured this way is the extinction by the ISM of the galaxy on 
the stellar light, $E(B-V)_s$, in which the geometry of the dust in the galaxy 
can play an important role (e.g. dustlanes as opposed to an homogeneous dust
distribution). 
A study of a large sample of hosts has been performed through HyperZ template fitting by Christensen et al. (2004), who find that GRB hosts generally exhibit little extinction, and have young stellar populations. The dependence on metallicity and assumed initial mass function is small (Gorosabel et al. 2003a,b; Christensen et al. 2004). One of the difficulties faced here is that the galaxies are often very faint.

\subsubsection{Emission-line spectroscopy}
In low redshift ($z \lesssim 1$) cases, an optical spectrum of the host galaxy can be taken,
typically showing several nebular and Balmer emission lines. The Balmer lines can be used to derive values for the reddening by calculating their deviation from case B recombination values expected in typical starforming region conditions (Osterbrock 1989). The derived reddening 
$E(B-V)_g$ is the reddening of the ionized gas in the source, i.e. the
dominant starforming region(s) producing the Balmer emission
lines. In general the reddening found from the Balmer decrement is low (see
e.g. Prochaska et al. 2004), to very low (Wiersema et al. 2006). 

The two host galaxy extinction estimates $E(B-V)_g$ and $E(B-V)_s$ may be correlated for
most galaxies (see Calzetti 2001 for a review). 
The increasing data volume on nearby GRB host galaxies will allow a test of
these correlations,
providing further insight into effective GRB host galaxy extinction curves.

On rare occasions it is possible to obtain high-resolution spectra of an
afterglow that also shows host galaxy emission lines, allowing one to obtain a simultaneous view of the extinction along the line of 
sight and of the Balmer decrement. In the case of GRB\,060218, both absorption
lines and emission lines are detected at high resolution using the UVES
spectrograph on the VLT
(Wiersema et al. 2006). The spectrum shows asymmetric emission lines that are
well fitted with two Gaussians separated by 22 km s$^{-1}$. The same two
velocity components can be seen in absorption in Ca II and Na I, and have
different chemical properties. These two systems can be interpreted as two
seperate starforming regions, through which the light of the afterglow
shines. A broadband measure of the extinction either from the afterglow or from
template fitting of the host would not have been able to seperate out the
contributions of the two individual systems.

\subsubsection{The longer wavelengths}
Yet another way to detect the presence of dust, is the detection of GRB host galaxies in the far-infrared
or sub-mm (see e.g. Barnard et al. 2003; Tanvir et al. 2004), where the UV radiation from massive stars is reprocessed by dust and reradiated in the far infrared. 
Detection of hosts in the far infrared as well as optical can severely constrain their SED (Le Floc'h et al. 2006), providing estimates for the unobscured star formation rate of GRB hosts. In a few cases values
up to hundreds of solar masses per year have been reported, while optical
indicators give much lower values, indicating a lot of dust-obscured star
formation. A different probe of unobscured star formation is the radio continuum flux, which is thought to be formed by synchrotron emission from accelerated electrons in supernova remnants and by free-free emission from H~II regions (Condon 1992). 
Berger et al. (2003) performed a survey of host galaxies at radio and sub-mm
wavelengths, finding that a significant fraction of GRB host galaxies have a
much higher radio-derived star formation rate than optical methods indicate,
pointing again to significant dust extincted star formation but based on only
a few detections.

\subsubsection{Our method compared}
All methods considered, SED fitting of the afterglow as described and carried
out in this
paper is the most broadband view of line-of-sight properties. Given the very
well known underlying continuum shape, the extinction curve can in principle be
extremely well modelled using this method. As discussed in Section 1, the effects of dust extinction
in the optical are better measured at higher redshifts whereas the X-ray
absorption is best measured at lower redshifts, so this method is most
reliable for some middle range of redshifts centred around $z=2$ which
approximately corresponds to the current
{\it Swift} median redshift (e.g. Jakobsson et al. 2006b). As with any line-of-sight method,
the measured columns may not be representative of the host galaxy as a whole,
so comparison with the integrated host galaxy methods is important. The general drawbacks of this method lie in the inability
to disentangle the locations of absorbers, both intervening (between us and
the host) and within
the host itself. Intervening absorbers may be identified using afterglow
spectroscopy (Ellison et al. 2006).
Distribution of absorbers within the host may be tackled from the point of view of searches for variability
in the absorption which may indicate destruction of dust or ionisation by the
GRB and/or afterglow, and comparison of the measured columns with known absorbers such
as Molecular clouds.
Campana et al. (2006a) find
that the X-ray absorptions of 17 bursts (including most of this sample) are
consistent with the GRBs lying along random sightlines towards Galactic-like
Molecular clouds. Our sample is too small to make such a statistical
comparison, but we note that the Campana et al. sample includes the results of
the Stratta et al. (2004) study of the GRBs presented here for which we obtain
largely consistent extinction values.

Dust destruction is of course impossible to measure in a single SED,
but may be seen with multiple epochs of data. 
Destruction of grains by for example the afterglow UV/X-ray emission
would in principle be observable, because depletion indicative line ratios
would vary on observable timescales as metals are released from grains into
the ISM. Variable line emission has been searched for in afterglows with
multiple optical spectra and has been seen in only two cases, GRB\,020813, in
which the {\small Fe II} $\lambda$2396 transition equivalent widths decreases
by at least a factor of five over 16 hours (Dessauges-Zavadsky et
al. 2006) and GRB\,060418 in which several transitions of {\small Fe II} and
{\small Ni II} are seen to vary in observations covering 11 to 71 minutes
post-burst (Vreeswijk et al. 2006). This line emission is thought to arise at 50-100 pc from the GRB
site, possibly within range of the GRB ionising flux. A variable $A_V$ has only been suggested for GRB\,980703 (described
in Section
4.1.6). There is evidence for time variable X-ray absorption in a small
number of GRBs ranging from tentative to moderately strong, in particular for
GRBs 011121 (Piro et al. 2005), 050730
(Starling et al.~2005) and 050904 (Bo{\"e}r et al. 2006; Campana et al. 2006b; Gendre et al. 2006) and 060729 (Grupe et al. 2006). This implies
ionisation of the line-of-sight gas, probably by the high energy GRB jet
(though we note that column density changes and early-time
spectral evolution can have similar effects on the spectrum).

Dust destruction can, however, only affect the immediate environment around a
burst, (perhaps out to 10 pc, Waxman \& Draine 2000), and not all the dust in
that region need be destroyed since this depends on the efficiency of the
mechanism and on the dust grain size distribution. Kann et al. (2006) noted
a possible tentative correlation between $A_{V}$ and star formation rate as measured via host galaxy
emission in the sub-mm (for a sample of 7 GRBs). If this turns out to hold
when tested against larger samples it would demonstrate that a substantial
amount of the dust we see comes from star forming regions located throughout the
whole host galaxy rather than very close to the burst where it could be
destroyed. However, this correlation is unlikely to be real given the very low
resolution of the sub-mm observations which make it difficult to differentiate
between host galaxy emission and field galaxy emission. This differentiation
can be made with Spitzer and lower star formation rates in comparison with the
sub-mm derived values have been found for some hosts including those of GRBs
980703 and 010222 (Le Floc'h et al. 2006). 

\section{Conclusions}
Here we have demonstrated the advantages of simultaneous fitting of broadband
data using a subsample of {\it BeppoSAX} GRB afterglows. In no cases is a
MW-like extinction preferred, when testing MW, LMC and SMC extinction
laws. The 2175\AA\ bump would in principle be detectable in all these
afterglows, but is not present in these data. An SMC-like gas-to-dust ratio
(or lower value) can be ruled out for 4 of the hosts analysed here (assuming SMC metallicity and extinction law) whilst the remainder of the sample have too large an error to discriminate.    

We discuss the various methods employed to derive host galaxy extinctions, and
compare our results with previous works. We find that this method provides similar central values of $E(B-V)$ and $N_{\rm H}$ to previous works in which extinction or absorption is determined through afterglow continuum fitting, and in the majority of cases we obtain tighter constraints.
We confirm that, with respect to continuum fitting methods such as this,
optical extinction is overestimated with the depletion pattern method, and
quantify this for a small number of cases.

Swift, robotic telescopes and Rapid Response Mode on large telescopes such as the William Herschel Telescope on La Palma and the Very Large Telescopes in Chile now allow earlier and higher quality data to be obtained, which will help immensely in discriminating between the different extinction laws at work in the host galaxies.

\acknowledgments
We are grateful to Mike Nowak for his assistance with {\small ISIS}. 
We thank Wim Hermsen, Jean in 't Zand, Erik Kuulkers, Tim Oosterbroek and
Nanda Rea for advice on the {\it BeppoSAX} data reduction, and the referee for
useful comments.
This research has made use of {\small SAXDAS} linearized and cleaned event
files produced at the {\it BeppoSAX} Science Data Center.
The authors acknowledge benefits from collaboration within the Research
Training Network `Gamma-Ray Bursts: An Enigma and a Tool', funded by the EU
under contract HPRN-CT-2002-00294. RLCS and ER acknowledge funding from PPARC.



\end{document}